\documentclass{report}
\begin{document}
\renewcommand{\thefootnote}{\roman{footnote}}
\title {Mathematics of Human Motion: from Animation towards Simulation (A View form the  Outside)}
\author{A.I. Zhmakin\\
ai@zhmakin.ru\\
Soft-Impact. Ltd, {Saint-Petersburg, 194156, Engelsa av. 27}}
\maketitle
\tableofcontents

\chapter*{Preface}

\begin{flushright}
{\small
And what is the use of a book without pictures or conversations? \par
{\em Lewis Carroll}
}
\vspace{.3cm}
\end{flushright}

Simulation of human motion is the subject of study in a number of
disciplines: Biomechanics,
Robotics, Computer Animation, Control Theory, Neurophysiology,
Medicine, Ergonomics.
Since the author has never visited any of these fields, this review is indeed
a passer-by's impression. On the other hand, he happens to be a human
(who occasionally is moving) and, as everybody else, rates himself an expert
in Applied Common Sense. Thus the author hopes that this view from the {\em
outside} will be of some interest not only for the strangers like himself, but for those who are
{\em inside} as well.

Two flaws of the text that follows are inevitable.
First, some essential issues that are too familar to the specialists to discuss
them  may be  missing.
Second, the author probably failed to provide the uniform "level-of-detail"
for this wide range of topics.

\chapter{Introduction}
\begin{flushright}
{\small
Before I saw the light and realized how evil physically-based
modeling is, I used to do it myself. \par
{\em James Blinn}
}
  \vspace{.3cm}
\end{flushright}

Computer animation is just a sequence of the objects' images displayed on a
screen. There are different classification systems for animation methods.
For example, Zeltzer's taxonomy \cite {zel85} includes guiding, animator-level
and task-level animation systems. In guiding systems, the behaviors of animated objects are explicitly
described while in animator level systems, they are algorithmically specified.
In task level systems, the behaviors of animated objects are specified in terms of events and
relationships.

Classification both according to the method of controlling motion and
according to the kinds of interactions the actors have has been proposed  by
Magnenat-Thalmann \&  Thalmann in \cite {tha91}. A motion control method
specifies how an actor is
animated and may be characterized in compliance with the type of information to which it is privileged
in animating the synthetic actor. For example, in a keyframe system for an articulated body,
the privileged information to be manipulated is the joint angles while in a forward dynamics-based system,
the privileged information is a set of forces and torques.
The nature of privileged information for the motion control of
actors falls into the three categories: geometric, physical and behavioral,
giving rise to three corresponding categories of motion control methods:
\begin{enumerate}
\item {Methods heavily relied
upon by the animator: performance animation, shape transformation, parametric keyframe animation.
Animated objects are locally controlled. Methods are normally driven by geometric data.}
\item {Physically-based animations,
especially dynamic simulation. The animator should provide physical data corresponding to the complete
definition of a motion. The motion is obtained by the dynamic equations of motion relating the forces,
torques, constraints and mass distribution for the objects. As trajectories and
velocities are obtained by solving the equations, motions are globally
controlled.} \item{Behavioral animation. Control of animation may be performed
at a task level or the animated objects can be considered as autonomous
creatures.} \end{enumerate}
The last one will probably became a mainstream animation method in the near future. The animation system
will accept natural language input and will store libraries of parameterized "self-knowledgeable"
characters that can understand the context of the environment and adjust their behaviors accordingly
\cite {bad00}.

At present, however, the animation methods are usually divided into three basic categories: keyframing,
motion capture and simulation.
Sometimes this taxonomy is extended by adding motion editing as a separate
approach \cite{gle98}.
Evidently, there is a trade-off between the level of control that the
animator has over the
fine details of the motion and the amount of work that the computer does automatically.

  \vspace{.3cm}

People are extremely skilled at perceiving the subtle details of human motion.
They are able to quickly pick out action that is unnatural or implausible
without necessarily knowing exactly what is wrong. Thus
humans in
computer animations  must be natural-looking
in both static postures and motion. It takes skilled animators huge amounts of time to produce
a quality animation of an articulated figure and integrate a character's
motion into a scene. Physically-based simulation is aimed at reducing both
time and a required skill of the animator while providing realistic motion.
 Forces associated  with the real world phenomena  (gravity, contact forces, torques on joints)
 are considered  along with the physical and physiological properties of the character (masses and
 inertia moments of the body parts, muscle parameters etc.) and the motion of the character
 is generated.  This is a computationally intensive process, involving the solution of many equations,
 first to
 determine accelerations for all the rigid bodies in the scene and then to produce motion over a
 time step.

  \vspace{.3cm}

Physically-based simulation of human motion can be used for the animation of human figures in
two
different modes:
editing (transformation) of the stored motions and generation of genially new
ones:

1. One could name a few actions that could be classified as the motion
editing: interpolation between motion captured frames, retargetting the
motion to a new character (or to the same one in another mood), adaptation to
the variations in the environment (changing the terrain profile,
for example), interpolation to produce a new motion by mixing
available in the library. The first of these examples is relevant to the keyframing technique
as well.

In all these applications the motion transformation could be performed either
in a pure geometrical way or be based on the underlying physical laws. The
motion transformation based on the geometrical consideration of the keyframe
images only (interpolation synthesis, "shape and animation by example", time
warping etc.) could result in artifacts such as a figure's distortion (a limb
shortening, for example).

2. Producing new animations from scratch could greatly simplify the
animator's work reducing it from tedious labour with great number of frames to
the script-like programming that specifies the goal of the motion and leaves
the generation of the low-level instructions to the system ({\em "George {\bf
goes} to the bar, {\bf takes out} a decanter and {\bf fills} the glasses"}).
This type of software is by far more difficult to develop.

The price for this "task-level" animation is, of course, the loss of the animator's control over
the subtle details of the motion.
However, animators usually don't want just a
pure simulation.
Animation and simulation have different nature:
aesthetics and control are of primary interest for the
first one (as has been pointed by Lee \cite{lee}, the results of simulations
are known before they begin). Correspondingly, the elements that constitute
animation  (modeling, animation, rendering - just mimicking
the traditional animation) and simulation (model creation, decision making, temporal management,
display building) are different \cite{lee}.

As was coined by A. Barr,
"physics by itself does what it wants to do -- it doesn't
want to do what you want to do". What animators really want is to
 guide  the models, to be able to specify some amount of
control over the motion and
then let the rest be automated.
Thus one needs both simulation and control to create animation
that does what one wants.

The optimal "mixing" of control and simulation in video games is formulated by Helava \cite{seppo}:
the {\em intended} motion such as running should be ("actively") controlled by the player
(and the response
time should be short enough to prevent her/him from wandering if the control
has been removed from her/his hands) while {\em unintended} motion (falling,
for example, or interaction with an obstacle)
should be generated by simulation
or physically based motion capture editing\footnote{Some form of "passive"
control should be still available to the player in certain circumstances (for
example, for  coordination of the body's limbs during a fall to make it
less painful).}.

\vspace{1.truecm}

The cardinal question is "What animation is {\em natural (believable,
convincing, realistic)} and are there any objective criteria?". Evidently, in
addition to the expert's judgment, one can try to quantitatively compare a
simulated motion
with either a digitized motion capture or with a simulation based on a {\bf
"full"}
model  using some appropriate norm to measure the deviation.
This leads to another question: what is a {\bf "full"} model and could it be
made
computationally tractable?

One can start with a model of the entire musculoskeletal system as a set of
{\em limbs}
considered as rigid bodies connecting  {\em joints} that, depending on their
nature, allow limb rotation around one, two or three axes (and gliding of the
limb for some joints) {\bf plus} a set of muscle-tendon actuators that
represent the
basic properties of muscles during force generation. The natural hierarchical
model allows one to compute a position of a particular limb by the
concatenation of the transformation in all joints along the path from the
root limb to the current one. The complete model of the musculoskeletal system
could be huge (there are about 150 mobile bones and over a hundred hinge,
prismatic or ball joints, thus a total number of degrees of freedom for the
skeleton alone without accounting for muscles or skin is about 250). However,
the model can be greatly reduced for certain types of motions.

One can consider different constraints on the human motion: {\em
anthropometric,
physiological, mechanical}.

Anthropometric constraints are formulated mainly as geometric ones. The most
common example is the range of admissible angles of the limb rotation in the
joint. It should be noted, however, that these limit angles determined by the
skeleton structure could be modified for physiological and psychological
reasons (for example, flexibility of muscles or existence of "comfortable"
limit angles).

Physiological constraints are related to the behavior of the muscle-tendon
actuators and could be specified, for example, as force-velocity relations
such as Hill law.

To simulate human motion dynamically, we should integrate the laws of
mechanics for all components of human model, and, therefore,  know all
forces and torques. Evidently, the wire-human model could not describe human
motion adequately. One should describe in addition the body's mass
distribution and the moments of inertia.

Unfortunately, it is not evident if these constraints are sufficient or
necessary
conditions for the motion to be believable. Evidently, the motion could
be anthropometrically valid and obey the laws of mechanics and
still look unnatural\footnote{A very simple example from classical mechanics: the falling
of an iron is identical to the falling of a bird's feather in vacuum. Still I am afraid that a lot
of people will consider such a motion as unbelievable.}. On the other hand, there is no reason to state that
physically impossible motion could not be considered as natural and that is
impossible to feign reality. Moreover, it is frequently possible (or even desirable) to
exaggerate\footnote{in fact, exaggeration is one of the basic principles of
the classical animation \cite{dis}} the physical effects and thus to produce
more "picturesque" motion. Brazel et.al. \cite{barz1} have suggested to
distingwish {\em physically} plausible and {\em visually} plausible motions.

The simulation task is complicated by the animator's desire to retain control over the motion and
to this purpose to invert the problem: rather
than specify forces and torques acting on the body and follow its motion,
she/he wants to get a motion from a given starting position to the final one
(inverse kinematic/dynamic technique). Thus some criterion to choose an {\em
optimal} trajectory is needed (minimum power consumption of all the
character's muscles, minimum displaced mass or some others).

An essential ingredient of the realistic  human motion is {\em
personification}. It could be based on the individual features of the
character such as her/his height, weight, some physical defect (lame leg, for
example), specific repeated action (adjustment of the glasses, rubbing a cheek
etc.). Probably one of the ways to implement this (as well as changing of the
character's mood) is via global control of the motion's perturbation coupled to
 the texture mapping as the main tool in achieving personification in {\em
appearance}\footnote{Luckily, the animation tools are not mature enough to bring into life a
nightmare of perfect digital copies of specific individuals forecasted by Badler \cite {bad00}
that inevitably will be drown
the Computer Graphics community in the marsh of legal and ethical problems.}.

\vspace{1.truecm}

The goals of the present study were:

A. To asses the state-of-the-art of the methods for simulation of human
motion for  animation of human figures as either an editing tool or an
automatic motion generator. The following  aspects have been  considered:

\begin{itemize}
\item{anatomical and physiological validity of human models}
\item{kinematic and dynamic human motion simulation methods }
\item{control over human motion}
\item{approaches to personification of human motion}
\item{motion editing  methods}

\end{itemize}

B. To suggest an approach to development of a hierarchy of mathematical models
and efficient simulation methods for producing realistic human motion.

\chapter{State-of-the-Art of Physically-Based Human Motion Animation}

\begin{flushright}
{\small
 The length of a progress report is inversely proportional to the amount of
progress.\par {\em More Murphy's Law}
}
\end{flushright}
 \vspace{.3cm}

 Keyframing requires that the animator specifies critical, or key, positions
for the objects.  The computer then fills in the missing frames by smoothly
interpolating between those positions.  This method requires the animator to
possess a detailed understanding of objects' motion  as well as a skill to
express that information through the keyframed configurations.  The continued
popularity of keyframing comes from the degree of control it allows over
the  fine details of the motion.

In  motion capture  magnetic or vision-based sensors record the actions of a human
subject in three dimensions. Afterwards stored data could be processed for animation of an
articulated figure.
This method is used widely since many commonplace human actions can be easily recorded.
However, it has a number of intrinsic limitations.
It is difficult to record certain movements: motion capture has the same
restrictions as live actors plus a few specific ones. Shifting of markers
placed on skin and clothing  as the human moves produces errors in
the data.
Magnetic systems often require the subject to be connected to a computer by
cables, restricting the range of motion. Optical systems have problems with
occlusion caused by one body part blocking another from view. Another problem
is difficulties  of  automatic  distinguishing the reflectors when
they get very close to each other during motion. These problems may be
minimized by adding more cameras.  And last but not the least, to be useful for
editing, the captured motion should be "projected" onto the articulated human
model (see a brief discussion of this problem below).

As a technique for generating human motion, physically-based simulation has two potential
advantages over keyframing and motion capture. Firstly, simulations can easily be used to produce
modified motion while maintaining its natural-looking character.
Secondly, real-time simulations allow interactivity, an important feature for virtual
environments and video games in which artificial characters must respond to the actions
of a player. In contrast, applications based on keyframing and motion capture select and
modify motions from a precomputed library of movements. This advantage is probably the most
important for animation of a transition from a specific motion to another one (for example, falling of
a running human).

A general hierarchy of the human models for animation has been proposed by Funge \cite{fun}:
 {\normalsize
 \begin{itemize}
\item{Geometric}
\item{Physical}
\item{Biomechanical (musculoskeletal)}
\item{Biomechanical (neuromuscular)}
\item{Behavior (low-level behavior)}
\item{Cognitive (high-level behavior)}
\end{itemize}
 }

Badler  \cite{bad1, bad2}
introduced a notion of {\em virtual fidelity} and listed five dimensions that could be used to
measure the quality of human animation:
{\normalsize
\begin{enumerate}
\item{ Appearance: 2D drawings $\Rightarrow$ 3D wireframe $\Rightarrow$ 3D polyhedra
$\Rightarrow$ curved surfaces $\Rightarrow$ freeform deformations $\Rightarrow$
 accurate surfaces $\Rightarrow$ muscles, fat $\Rightarrow$ biomechanics
 $\Rightarrow$ clothing, equipment $\Rightarrow$ physiological effects
 (perspiration, irritation, injury)}
 \item {Function: cartoon $\Rightarrow$ jointed skeleton $\Rightarrow$ joint limits
 $\Rightarrow$ strength limits $\Rightarrow$ fatigue $\Rightarrow$ hazards
 $\Rightarrow$ injury $\Rightarrow$
 skills $\Rightarrow$ effects of loads and stressors $\Rightarrow$ psychological models
 $\Rightarrow$ cognitive models $\Rightarrow$ roles $\Rightarrow$ teaming}
 \item {Time: off-line animation $\Rightarrow$ interactive manipulation $\Rightarrow$ real-time
 motion playback $\Rightarrow$ parameterized motion
 synthesis $\Rightarrow$ multiple agents $\Rightarrow$ crowds $\Rightarrow$ coordinated teams}
 \item {Autonomy: drawing $\Rightarrow$ scripting $\Rightarrow$ interacting $\Rightarrow$ reacting
 $\Rightarrow$ making decisions $\Rightarrow$ communicating $\Rightarrow$
 intending $\Rightarrow$ taking initiative $\Rightarrow$ leading}
 \item {Individuality: generic character $\Rightarrow$ hand-crafted character $\Rightarrow$
 cultural distinctions $\Rightarrow$ personality $\Rightarrow$ psychological-physiological
 profiles $\Rightarrow$ gender and age $\Rightarrow$
 specific individual}
\end{enumerate}
}
Evidently, the relative importance of these parameters depends on the application.

\section{Anatomically Correct Human Figure Presentation}
\begin{flushright}
{\small
The head sublime, the heart pathos, the genitals beauty, hands and feet
proportion \par {\em William Blake}
}
\end{flushright}
\vspace{.3cm}

The basis of believable human motion is a correct representation
of the human figure. One can distinguish two aspects of a realistic human model: its resemblance to
the anatomically plausible structure (first of all in the topological sense) and the richness of a model in details
(human figures
have been animated using a variety of geometric models including stick
figures, polygonal models, and NURBS-based models with muscles, flexible skin,
or clothing). It is interesting to note that while the simulated motion does not look as natural as
that of real people, for the time being,
some viewers do choose simulated motion as more natural when both types are viewed with the
human bodies removed so that the movement is shown only as the dots located at
the joints.

Evidently, blunt errors in correspondence between
articulated skeleton and real human one should more explicitly manifest
themselves in motion   than in static postures. As an example, animation of the
golf player described by Henry-Biskup \cite{henry-b} could be mentioned. The model was
attached  to a rather simple articulated skeleton driven by motion captured
data. It is turned out that the golfer posture during bending is very stiff and unnatural,
thighs seem to be too long and the shoulders ballooned out. All the attempts to
improve appearance by changing vertex attachments and editing link parameters
failed. The problem was solved by examination of the underlying skeleton
structure and  its modification. First of all,  a coplanar arrangement of two hips
and back joints on one horizontal plane was changed to more anatomically
motivated one with lowered hip joints. The second necessary modification was
raising of the shoulder joint.

Another example is the spine modeling.
Spinal column consists of 24 movable vertebrae \cite {beg}. The accumulation of each
vertebrae's small rotations results in considerable bending and twisting.
The spine connects  three large bony masses of the skeleton - the skull,
rib cage, and the pelvis. Ratner \cite{rat} stresses that generating realistic motion required an adequate
description of the action of the spinal column and its effect on the torso.
General aspects of the biomechanically sound approach to the human motion have been discussed by
Zatsiorsky et.al. \cite{zat0}.

It should be noted that due to the hierarchical
structure of the human figure the error is more dangerous the closer it
is located to the root.

An advanced model of an articulated figure based on biomechanical data is described by Savenko et.al.
\cite{sav}.
The authors state that the usual hierarchical model of articulated figures in animation
assumes that all joints having more than one degree of freedom
can be effectively represented as several hinge joints connected by zero-length links.
Thus the motion in such a joint can be modelled as two or three successive rotations
about axes of the fixed orthogonal reference frame.

This type of model is highly suited for robotics, where the joints are
carefully constructed to have a prescribed number of degrees of freedom, but
this does not guarantee that it is appropriate for figure animation. It is
obvious that the structures of real joints are quite different from those of
mechanical joints.

The following assumptions are usually made in the models of an articulated figure:
\begin{itemize}
\item{the axes of the joints are parallel to the anatomical axes of the figure,
i.e. all basic motions in a joint with one degree of freedom are performed in
either the frontal or the sagittal (vertical plane which divides a body into
two symmetric parts) or the transverse (horizontal) planes}
\item {the axis of rotation of a joint does not change during the motion}
\end{itemize}

However, for some joints the inclination of the axis in one or two planes can be as large as
up to $40^o$  \cite{sav}.

A joint is a bodily component used to connect the bones of the skeleton.
Apart from connectivity, joints also provide mobility and stability.
The structure and function of joints are closely related. The structure defines
the character and range of motion available to the joint.   A joint capsule
encloses the ends of the bones and thus connects them, but not in a rigid manner.
In order to obtain additional stability and strength for a joint, the bones
are also connected by ligaments.

Most joints in the human body permit some form of motion and  are divided into
three groups: uniaxial, biaxial and multiaxial on the basis of the motions which can take place.
This classification is simplified; most joints show supplementary motions
(both rotational and translational) of small amplitudes\footnote{There is an opinion that inability of the
joint prostheses to allow such motions is one of the main reasons of their long-term failure \cite{vans}.}.
These motions, called combined, occur unconsciously, with little (if
any) participation of the central nervous system \cite{vans}. The joint
surfaces and ligaments are playing the dominant role. For example, the
internal rotation of the lower leg associated with the knee flexion (with an
amplitude about $15^{\circ}$) is controlled by the knee ligaments and asymmetry of
the knee joint surface morphology. However, the primary motion of the knee
flexion is performed voluntarily.

The axis of rotation does not stay the same, but continually changes throughout the motion \cite{for,sav}.
However, a single stationary axis is a reasonable approximation for many joints.
The idea of an {\em instant} axis of rotation can be used to model
complex simultaneous motions in a multiaxial joint. For instance,
a motion in the hip can be modelled either as a sequence of three rotations
about three axes or as a single motion about the instant helical axis.

Sometimes a combined motion can occur in the joint. For example,
a coupled rolling-sliding motion is observed in a human knee during flexion
\cite{lu}. Evidently, the most accurate (but computationally intensive) method
to study joint motion is numerical simulation of a free joint (no fixed
restriction on the degrees of freedom) using finite element representation of
the bones in contact \cite{for}. Effective restriction of the degrees of
freedom will be provided by contact forces between bones and by ligaments and
muscles.

A simplification  of the anatomical structure is used often when gross motion
is simulated. For example, the shoulder complex, which refers to the
combination of the glenohumeral joint and the shoulder girdle including
the clavicle and scapula and their articulations, has been studied in detail in
\cite{lind}. The model contains four bones (thorax, clavicle, scapula and
humerus) together with four articulating joints. Twenty-two muscles serve the
shoulder mechanism and provide the variety of motions of the human upper
limb. The greater part of these muscles are broad and have no unique
line of action. Thus for modeling these muscles with large attachment areas
splitting  into several elements should be used. In the cited paper thirty-six
elements have been considered as the stretched strings and the muscle forces
have been assumed to act along these direct lines. It should be noted,
however, that in reality many of the muscles follow the contours of the body
structures and, therefore, would be better modelled as the curved elements
\cite{helm}. Nevertheless, such complex structure as the shoulder girdle
could be reduced to
a "lumped" spherical joint between thorax and humerus \cite{wang} when the
gross human
motion is of interest.

A description of the general aspects of the human joint geometry and
kinematics as well as information on the twenty-six specific human joints
can be found in a book by Zatsiorsky \cite{zat}. A detailed modeling of
the upper human limb has been performed by Maurel \cite{maur1,maur2}.

The skeleton being defined, one has to divide human body into segments
(which is not a trivial task as explained in detail by Hatze \cite{hatze}) and
to determine parameters of these segments. Among these parameters mass,
volume, density, position of the centre of mass and moments of inertia
should be mentioned. Bj\/ornstrup \cite{bjor}  reviewed the methods for
measuring of the body segment parameters starting from probably the
first registered work by Borelli (1680). He also gives references to a number
of published papers that contain measured results for different parts and
for the whole human body.

Kepple et. al. \cite{kepp} have examined the muscle attachments from 52 skeletons
and suggested the locations of idealized muscle attachments on the
             pelvis, femurs, tibiae and fibulae, and feet.
             Statistical accumulation and scaling techniques have been used to
             generate highly representative normative models, which have been
             divided into groups and tested for differences based on gender
             and race. From the test results, the pelvis was divided into a
             male model, a black female model and
             a white female model. The foot was separated into
             black and white models. Single
             models were used for the femur and the
             tibia/fibula.

\vspace{.3cm}
A requirement for the complexity of a universal human articulated figure has
been suggested by Boulic \& Mas \cite{bmas2}. The authors assume that one
needs at least 30 DOFs for extremities alone plus 12 DOFs for a crude
representation of the relative motion of the pelvis, abdomen,  thorax,
 neck and  head. Thus, even excluding the mobility of the clavicle and
scapula, the model should contain 42 DOFs. Additional 40 DOFs are needed for a
simplified description of the hands. Adding 8 to 12 DOFs for animation
of the clavicle and scapula and a few DOFs for "aggregate vertebrae" to
provide spine bending one would get over 120 DOFs depending on the resolution
of the spine model.

This
model  is in close agreement with that of Hatze \cite{hatze} who assumes that for
the simulation of gross body motions a 17-segment figure is optimal
(abdomino-pelvic segment, abdomino-thoraic segment, head-neck, left and right
shoulders, arms, forearms, hands, thighs, legs and feet). The model has 42
DOFs (3 linear and 39 angular generalized coordinates).



\section {Neurophysiological Aspects of Human Motion}
{\small
\begin{verse}
If you can force your heart and nerve and sinew \par
 To serve your turn long after they are gone    \par
\end{verse}
\begin{flushright}
{\em Rudyard Kipling}
\end{flushright}
}
\vspace{.3cm}

Humans move easily with little concsious thought.
This ease, however, is deceptive. It is a wonder, given the complexity of the
musculotendon system and of control processes that coordinate muscles
activity,
that a human can move at all\footnote{remember a malignant snail that asked a
centipede about the order it moves its legs in}.

The systematic analysis of the control of human motion is certainly beyond
the present review (see, for example \cite{bbm} and the references therein).
In what follows a few not-too-well-structured remarks are given.

To develop an adequate biology-based model one has to know the structure and
properties of the neuromuscular system and to understand the principles of its
operation. Van der Helm \cite{helm1} has listed the needed data:

\begin{itemize}
\item{the linkage system (bones, joints, ligaments) with the assotiated mass
and viscoelastic properties}
\item{the actuator system containing the muscles}
\item{the sensory system with the muscle spindles and Golgi tendon organs}
\item{the control system at various levels of the central nervous system}
\end{itemize}

A primary role of muscles is to produce forces as a result of contraction and
thus provide joint torques that cause limb movements.
Evidently, correct description of the muscle properties is essential
for simulation of human motion.

Muscle modeling should include two steps:
\begin{enumerate}
\item{determination of the muscle line of action and/or muscle moment arm}
\item{simulation of the muscle dynamics}
\end{enumerate}

There are a number of different definition of the line of action.
The two obvious
ones is straight line and centroid line approaches. Maurel \cite{maur1}
suggested a {\em contour} line method and van der Helm \cite{helm1} - a {\em
bony contour} approach. Muscles with broad attachements should be divided into
several bundles.

The force produced by muscle is
determined by its size and structure, the contractile conditions (length and
speed of contraction) and the character of activation. The force is usually
assumed to be proportional to the physiological cross sectional area (PCSA).

The dependeces of the muscle force on the level of activation, length, velocity
have been studied by numerous researchers \cite{herz}.
The best known one is by all means Hill's  force--velocity
relationship \cite{hill}.
It should be noted, however, that
is is valid for maximally activated muscle, isotonic shortening contractions
and at the optimal length of the muscle. The fact that the force--velocity
relationship is not an intrinsic muscle characteristic has been recently
confirmed by Ettema and Meijzer \cite {ette}. The authors have compared a
modified Hill's model and an "exponentially decay model" and conclude
that  both well describe short contractions (300--500 ms) and less accurately
long duration behavior. Since exponentially decay model does not incorporate
a force--velocity curve, the authors argue that this relation is not a
fundamental
property of muscle in contrast to, for example, the lenth--tension curve.

Phenomenological Hill's model that includes a contractile element, a series
elastic element and a parallel elastic elements or its modifications are used
in the majority of works. The relation of the muscle properties (active and
passivwe length --tension, isotonic force--velocity) to the anatomical
structure of muscles has been studied by Lieber \cite{lieb}. A comparison
of the Hill model with the distributed moment model based on the cross-bridge
theory
of muscle contraction for iso-velocity stretches performed
by Cole et.al. \cite{cole} have shown that the former model is preferable in the
cases considered. A model accounting for the details of muscle activation
(time pattern of the stimulation interpulse
interval, pulse width and pulse amplitude) has been developed by Riener \&
Quintern \cite{rien}. Three anatomy-based muscle models have been constructed
and applied to the muscles of the arm and torso by Scheepers et.al.
\cite{schee}.

 A thermodynamic analysis of the muscle contraction in which muscle, not an individual
 myosin crossbridge, is considered as a near-equilibrium system has been applied by Baker \& Thomas
 \cite{bake}
 to a simple two-states cross-bridge scheme \cite{huxl}. As a result the authors
have been able  to provide a theoretical basis for an empirical Hill's equation.

An advanced  computational approach to modeling the complex mechanical
properties of muscles and tendons under physiological conditions of
recruitment and kinematics has been developed by Cheng et.al. \cite{che}. The model is based on recently published
data on muscle and tendon properties measured in feline slow- and fast-twitch
muscle, and incorporates a novel approach to simulating recruitment and
frequency modulation of different fiber-types in mixed muscles.

It should be noted that practically all the papers devoted to the muscle contraction simulation
 (based on either simple Hill-type models or more elaborate ones that account for excitation
 dynamics) assume muscle uniformity. Thus it is possible to consider "lumped" model consisting of
 a few zero-dimensional elements. In reality, muscle is far from being uniform,
especially in cross-sectional area  and activation pattern. Its contraction
should also be nonuniform as has been  proved by continuum one-dimensional
computations performed by Aigner \& Heegaard \cite{aign}.

Accurate simulation of the tendon is almost as critical as that of muscles
themselves.
Tendon compliance is important because it causes the length and
velocity of the contractile element to be out of phase with those of the
whole muscle. For example, in an active muscle with a high ratio of
tendon-to-contractile element length, a stretch to the whole musculotendon
system would only slightly lengthen the contractile element, with the majority
of the length change being accommodated by lengthening of the tendon
\cite{zajac}.
In fact, under certain conditions, the musculotendon
element can be lengthening while the contractile element within the muscle is
shortening, or vice-versa. This behavior would be misrepresented in
tendon-less models, which assume that contractile length changes are
proportional to path length changes.
 The tendon and the
contractile muscle fiber elements both act on a muscle mass that provides
inertial damping to prevent instabilities  arising within
the muscle model \cite{loeb}.

A Hill type model model that accounts for nonlinear static and dynamic
properties of both muscle and tendon have been used by Riener et.al
\cite{rien1} to develop a model of the human knee. Muscle fatigue and passive
muscle viscosity have been incorpoarted into the model.

Zajac \cite{zajac} has advocated a model that defines the properties of a
generic musculotendon element with a single set of equations.  The
five parameters required by Zajac's equations are optimal fascicle
length (the length at which the muscle produces maximal tetanic isometric
force), maximal tetanic isometric force, pennation angle of the
muscle fibers at optimal fascicle
length, time-scaling parameter for maximal muscle
shortening velocity and rise and fall times, and tendon slack length.
Once these measures had been determined for a given muscle, its behavior could
be reproduced mathematically.

A special attention should be payd to ligaments. The high
stiffness of ligaments result in a high sensitivity of the ligament forces to
the joint motions. For the inverse dynamic simulations it is virtually
impossible to include
ligaments, in forward dynamic ones the stiffness of ligaments is a major
factor determining the size of the time step \cite{helm1}.

\vspace{.3cm}

The most unconvenient feature of the musculoskeletal system for the numerical
simulation is by all means {\em redundancy}. One should distingwish
an actuator and kinematic redundancies. The first one is in that more than
one muscle contols one degree of freedom (DOF). For example, for a simple
hinge joint with 1 DOF there are two muscles (agonist--antagonist).
Muscle can co-contract: while the net moment at the joint could be constant,
the forces of the muscles change.
Kinematic redundancy means that there are more DOFs in the joints than at the
end-effectors (for example, arm can move when a hand is holding
a handle).

The aim of the study of motor control is  to determine how the central
nervous system effects functional, goal-directed movement \cite{stei}.
A central issue for understanding human motion is how the excessive degrees of freedom
of the neuromuscular system are controlled \cite{Ber}.
Usually
the mechanisms involved in motor control are presented as a hierarchy, with the
brain at the top, the spinal cord in the middle, and the musculoskeletal
system at the bottom \cite{loeb}.

The importance of the muscles simulation for the correct description of the
movements themselves is evident.
However, it seems that their role is not restricted to the producing force.
Full \&  Koditschek \cite{fullk} have studied the significance of the neural
and mechanical systems in the locomotion control. The conclusion is that the
slow, variable-frequency motion is dominated by the nervous system while the
role of the mechanical system is significant in the control of the rapid,
rhythmic movements. As has been noted by Raitbert \& Hodgins,
"the mechanical system has a mind of its own"  \cite{raibh}.

Wagner \& Blickhan \cite{wagn} have studied stabilizing properties of skeletal muscles, i.e.
their ability to damp the oscillatory movement ("preflex") without reflexive changes in activation.
Four muscles model have been considered: "Hill--simple" (just the contractile element), "Hill--SEG"
(an additional serial-elastic element), "Hill--PEC-DC" (an elastic element and a damper connected in parallel
to the contractile element), "Hill--$E(t,X,V)$" (the contractile element is activated by the function
 representing a displacement and velocity sensitive controller).
 The authors conclude that self-stabilizing ability of the muskuloskeletal system is significant only if
 the muscle properties (force--velocity and force--length relationships) are tuned to the geometric properties
 of the linkage system.

One the most remarkable features of the human motion is the synergies in
operation.
 Synergy is the working groups of neurons, muscles,
joints as if they were a single entity.  This notion was introduced  by Bernstein \cite{Ber} to cope with the problem of
muscle and kinetic redundancy. An example of synergy is a linear relation between elbow and shoulder
dynamic torque in natural pointing movements \cite {zaal}.

Locomotion involves repetitive movements and is often executed
             unconsciously and automatically. In order to achieve smooth
             locomotion, the coordination of the rhythms of all muscles is necessary.
             Neurophysiological studies have revealed that
             basic rhythms are produced in the spinal network called the
             central pattern generator (CPG), where some neural oscillators
             interact to self-organize coordinated rhythms.

              Review of the experimental evidence for existence of CPG
              in cats and primates (including man) can be found in \cite{duys}.

              Recent experimental results indicate that decerebrate
             cats have the ability to learn new gait patterns in a changed
             environment\footnote{However, this does not include situations when
              the motor cortex is involved in anticipatory gait
           changes due to visual input (decerebrate cats cannot step over
           obstacles)\cite{taga1}} \cite{itos}.

            In the last years it has become possible to regain some
             locomotor activity in patients suffering from an incomplete
             spinal cord injury  through intense training on a treadmill.
             Non-patterned electrical stimulation  in subjects with complete,
             long-standing spinal cord injury, can induce patterned,
             locomotor-like activity \cite{dimi}.
              This finding suggests that spinal circuitry in humans has the
             capability of generating locomotor-like activity even when
             isolated from brain control.

             The evidence for the low-level muscle control by primary motor cortex
             and resolution of the seeming contradictions with experimental data
             is discussed by Todorov \cite{todo}.

              There is increasing agreement that the cerebellum plays
             an important role in motor learning \cite{smi}.
             A unique
             characteristic of motor learning is that it adjusts joint and
             limb mechanics by altering the neural input to muscles through
             practice and mental rehearsal. Smith \cite{smi} proposed a hypothesis
             that the cerebellum plays an important role in motor learning by
             forming and storing associated muscle activation patterns for the
             time-varying control of limb mechanics.  Optimal control
             cannot be achieved by online corrections initiated by reflex
             feedback because of the delays and consequent instabilities
             incurred.

              A discussion is under way about the use of
             representations (internal models) to explain how intelligent behavior is generated.
             These models are used by
             the subject to instruct the motor apparatus.
             So-called interactionists do not accept
             the existence  of internal representational models. One of the  problems
              non-representational approach faces is a difficulty with explanation
of the               anticipatory, future oriented behavior. Keijzer
\cite{keij} has made an attempt to               extend the interactionist
conceptual framework by               the analysis of
             anticipatory behavior as a process which involves multiple
             spatio-temporal scales of neural, bodily and environmental
             dynamics.

However, the available data strongly support the existence of internal models
for motor control and trajectory planning \cite {kaw}.  Moreover, it it is
shown that cerebellum should contains internal models of different types
\cite{kaw1}.
Forward internal models predict the consequences of actions and could overcome
time delays associated with feedback control.
Multiple paired forward and inverse models play an important role in motor
learning.

Human ability to imitate is not well understood but a powerful form of motor skill learning.
Evidence from neuroscience indicate that there are two neural structures that are of great
 importance to imitation.
The first one is {\em spinal fields} that contain code for the primitive motions \cite{bizzi}.
(i.e. stereotypical movements invariant to the exact position and the rate of motion).
Primitives can be combined to produce a meaningfull motion. The second one is that appear to directly connect the visual and
 motor control  systems by mapping observed motions to the motor structures \cite{rizz}{\em mirror neurons}\footnote{
mirror neurons have been shown to fire both when the monkey grasps a banana and when it observes
another monkey or a human performing a similar grasp}.

These findings have been exploited by Billard \& Matari\'c \cite{billa} for the development of the
imitation based control system for humanoids robots. The motor control in this system is
hierarchical with a spinal form module at the lowest level and cerebellum modules responsible for
the motion learning are at the highest one.

A mathematical model of the spinal motor control system (SMCS) has been
developed by Shimansky \cite{shim}. The author has shown that  an
internal
representation of the controlled object exists already at the level of SMCS. This
system is able to generate motor patterns for reflex rhythmic motions
such as locomotion without the aid of the peripheral afferent feedback
and can modify its activity in responce to peripheral afferent stimuli.

Hatze \cite{hatz}
has considered the myoskeletal and myocybernetic control problems
and have shown that, while both of them ill-posed, the latter is unsolvable in the present formulation ("find
neural control
that generate observed motion").
This is explained by the hyposensitivity of the skeletal movements to
the neural control input perturbations.

\section {Basic Approaches to Animation of Articulated Figures}
\begin{flushright}
{\small
 Man will occasionally stumble over the truth, but most of the time he will pick himself up and
 continue on.  \par
{\em Winston Churchill}
}
\end{flushright}
\vspace{.3cm}

    The literature on computer animation is vast (for example, a list of references
published by Magnenat-Thalmann \& Thalmann  in 1992 \cite{tha92} contained 600 entries) and its comprehensive review is by all means
beyond the present report. We are going to focus on the issues closely related to those approaches
to human motion editing/synthesis that rely heavily on biological knowledge and physical laws.
Moreover, we will not touch at all such topics as texture mapping and face animation \cite{face1,face},
and cloth \cite{cloth,cloth1,cloth2, pbw} modeling. The review,
being restricted in such a way, still will be certainly incomplete.
Some sections will inevitably be sketchy.
We hope to outline basic trends and approaches only and illustrate them
by a few typical examples, not bothering with establishing priorities
of different researchers.
The division of the reseach topics between subsections is rather
arbitrary (for example, motion editing can incorporate elements
of physically-based simulation; it is unclear what subsection the studies of free
form deformations or agent's path planning belong to).
Mathematical details  will be avoided wherever possible since they
are, as a rule, not animation specific and can be found in standard references.

\subsection {Kinematics Approaches}
 \begin{flushright}
{\small
As speed increases objects can be in several places at once.\par
{\em Cartoon Laws of Physics}
}
\end{flushright}

\vspace{.3cm}
 In kinematics, an articulated figure is a set of rigid bodies whose motions are
restricted by the linkages among them. The state of the system can be defined as a union of the states
of the bodies; the state of the body is just its position and orientation. Linkages (joints) are reducing the
total number of degrees of freedom (DOFs) of the system. In practice the figure is usually described in
an hierarchical way \cite{brud} and is organized in a form of a tree; the nodes
of the tree correspond to the links, and the edges correspond to the joints.
The purpose of the joint connecting two links is to perform a transformation of
one link relative to another. Thus the absolute position of a particular link
can be calculated by concatenation of the transformations in all joints along
the path from the root link to the current link. Thus the state of the figure
is given by the position of its root and the joints angles (one to three,
depending on the joint type) \cite{cal,pbw}.

The figure's motion is a rather rich problem for mathematicians. While the motion of a single rigid body
is well understood  (see the book by Arnold \cite{arn}), the
theory of multibody dynamics is still under development (for example, a
recent paper by Park \& Kim \cite{park} where coupling of the Lie group
techniques and Riemannian geometry has been exploited to study metric
characteristics of a generalized inertia tensor of the system and  approaches
to its factorization  for both open and closed kinematic chains\footnote{Note,
however, that \^Zefran et.al. \cite{zefr} have shown that no Riemannian metric
exists whose geodesics (a generalization of straight lines in Euclidean space
to Riemannian manifolds) are screw motions between two positions of a rigid
body and semi-Riemannian metric is more appropriate for description of such
motions.})).

There are different approaches to mathematical description of the figure
articulations. Matrix method allows one to define both the translation and the rotation of the local system
fixed to a body and, thus, the position of the object in the different systems
and its displacement can easily be defined. The components of the {\em
rotation matrix} are {\em direction cosines} of the local system relative to
the global one (direction angles are the angles that a vector makes with  the
coordinate axes). Combining translation and rotation one gets the  {\em
transformation matrix}.  The body motion can be presented as a sequence of the
rotations and translations.

 Nine direction cosines are, evidently, redundant, since the orientation of the body
 in space is determined by a smaller number of parameters. Three independent angles
 corresponding to the three rotational degrees of freedom ({\em Euler angles})                                                                             need to be determined
 \footnote{Euler theorem states that "Any motion of a rigid body with one point fixed is a single rotation around an
 axis through that point".}.
  Rotations in 3D space
 are noncommutative: changing their order will change the final body position.
 A specific order of the Euler angles is usually used in biomechanics \cite{zat}: precession
 (in the sagittal plane), nutation (away or towards the sagittal plane) and twist, or spin (along
 the axis fixed within a body). Sometimes the so-called {\em nautical} angles (yaw, pitch, roll) are used.
In total, there are 12 different combinations of rotations that can produce a given spatial orientation.

For the particular angular positions a singularity ("gimbal-lock") occurs: two
axes become parallel and Euler angles could not be determined (effectively,
one degree of freedom is lost). Another drawback of using Euler angles is the
jerky, unnatural motion produced when one interpolates Euler angles between
different positions \cite{ariel}.

The screw (helical) method based on the Chasles' theorem ("Any motion of a rigid body can be
obtained as the rotation around an axis  and the translation along this axis") permits a description of the
body motion without referring to arbitrarily chosen axes of rotation \cite{zat}.
At any given instant there exists a line ({\em screw}, or {\em helical} line\footnote{helical line does not
need to lie within the body}) that keeps its position in space and the translation and rotation
occur along and around this line. The following six parameters are required to define the body position:

\begin{itemize}
\item{two coordinates of the piercing point of the helical axis with one of three coordinate planes}
\item{two direction cosines of the helical axis}
\item{ body translation along the helical axis}
\item{ body rotation about the helical axis}
\end{itemize}

When helical axis method is used to describe rotations, only three parameters are needed.
One can use either four Euler' parameters subjected to a constraint equation, Rodriguez's
parameters or quaternions. Formally, the latter can be written as a sum of a scalar and a vector with
Euler's parameters as coefficients.

Quaternions are assumed to be superior over traditional matrix methods for
intensive computations with 3D rotations \cite{ariel}.
Hestens \cite{hest1} advocates the use of quaternions in the frame of {\em geometric algebra}
developed by the author . This description is invariant
(meaning coordinate-free) and, as the author argues, facilitates the analysis
of different control variables and kinematic constraints.

In fact, geometric
algebra is just another kind of symbolic
notation. Its main advantage over the classical vector and tensor
notation is the simplicity of the formulae and absence of numerous
indices typical for the tensor calculus. However, this notation is
not a new and cutting-edge technique. It is based on the quaternion
calculus and just applies the classical duality theory to this
finite-dimensional space. Invariance of this approach is very
useful for symbolic manipulations.
It is claimed in the cited paper, that geometric algebra is
computationally superior in comparison with the classical vector and
tensor notations. But computations themselves are based on the
Frenet's basis notation, not the geometric algebra. Probably
this claim means that if one uses geometric algebra for the
symbolic manipulation, the formulae obtained are simpler (even in
the terms of spherical trigonometry) than ones obtained with
tensor calculus. This statement is perhaps true, because the main
advantage of the geometric algebra as a symbolic language is
existence of so called {\it spinors} that define rotational
transformations. Such objects, combined with purely algebraic
technique (reduction of similar terms, factorization and so on),
really simplify both forms of the obtained expressions (invariant,
i.e. formulated in the dual space and covariant, i.e. formulated
in the original 3D space).

In the appendix to the second part \cite{hest1} a generalization of the
spinors concept to represent translation (hyperspinors) is proposed. Thus
Hestens is able to reduce the composition of arbitrary displacements to
multiplications and to obtain a universal description of evolution of a
kinematic chain via both rotations and translations.

\subsubsection {Forward Kinematics}
 \begin{flushright}
{\small
 You can't fall off the floor.\par
{\em Murphy's Law}
}
\end{flushright}
\vspace{.3cm}

It is natural to consider forward kinematics methods together with
procedural animation. The essence of these approaches is  the hierarchical
determination of the limb positions on the basis of given variations in time
of the joints angles.
In other words, they provide transformation of position and velocity from the
joint space (joint angles) to Cartesian coordinates.

  Forward kinematics approach provides a complete control
of the motion at very low computational cost $O(n)$, where n is the number of
the joints. Its advantage is the use of high level parameters (speed, step
length etc.) that allows the animator to easily generate families of different
motions.

There are two main drawbacks of these methods.
Firstly, special handling is required to satisfy obvious constraints imposed on the motion
(such as, for example, a feet contact with the surface during walking). In this example a popular
solution to provide that the supporting foot would not go through or off the ground
is to shift the root of the articulated figure to this foot.
Secondly, it is quite animator's time consuming and needs considerable skill to get convincing motion
and these requirements grow with increasing of the complexity of either the human model (number of
DOFs) or the motion.

These approaches are mainly useful for the well known movements whose
peculiarities are known from observations \cite {bru,gm85,zel82}.

\subsubsection {Inverse Kinematics}

While in a classical keyframe approach an animator has to directly
manipulate degrees of freedom of an object \cite {las87},
application of the inverse kinematics methods allows one to indicate
the position of specific points  of the articulated figure  ({\em end-effectors})
(such as a hand or foot) and let the computer to determine the values of the joints
angles that provide the desired configuration \cite {bmw,gm85}.
In other words, the end-effector position and orientation control all the joints
along an articulated kinematic chain \cite{kog,tol,zhao} thus reducing greatly the number of control parameters.
It should be noted, however, that the reconstruction of the intermediate joints angles is not unique
and, as was noted by Bernstein \cite{Ber}, the simplicity of the
end-effector trajectory does not imply that those of the intermediate joints
will be also simple (in fact, quite the opposite is observed).

Thus in addition to the movement of the end-effectors the animator should define a set of constraints
that will drive the limbs in a unique way. Some of constraints (mainly of geometrical nature:
feet position during locomotion, obstacle avoidance etc.) are obvious while
the others could be used to influence the style of motion \cite{bou1,pbw}.

A number of methods to implement inverse
kinematics approaches has been proposed. Zhao \& Badler \cite{zhao} have given
 a detailed description of the non-linear
programming algorithm that remains robust even for the articulated figures with a large number of
joints. All spatial constraints are divided into those that applied to the figure itself
and those that refer to the environment. Two approaches (the pseudo inverse method with
explicit optimization and the extended Jacobian method) have been studied
by Tevatia \& Schaal \cite{tev}.
The methods have been compared for the test case of a 10-DOF figure model. Results
for a 30-DOF model have been tested on a humanoid robot. Miller et.al. \cite
{mill}  proposed to use the singular value decomposition method for solution
of the underdetermined system of equations.

A variant of the inverse kinematics method called "differential inverse
kinematics"
is proposed by Chaffin  et.al. \cite{chaf}. The method is based on using velocities (rates)
instead of positions.

\vspace{.3cm}

 Boulic \& Thalmann \cite{bt} have demonstrated how one could combine direct
and inverse kinematics:
 a "leg-correction" procedure is invoked
if the foot penetrates the ground after forward kinematics
computation and inverse kinematic algorithm
is applied to modify the foot position.
The complexity of inverse kinematic algorithm ($O(n^3)$ due to the need to
invert the Jacobian matrix) is not a problem in this case since
it is usually used for one leg only.

\vspace{.3cm}

A possibility to extend the problem formulation and allow the animator to control not end-effectors
only but in addition the  centre of mass of the figure  and moments of inertia has been studied by
Baerlocher \& Boulic \cite{bae1}.  The control of COM is
needed, for example, to ensure the static balance of the complex articulated figure.

Integration of the mass distribution information to embody the position control
of COM\footnote{The importance of the COM control confirm recent experimental study \cite{pat}.
It has been shown that control of COM precedes all other changes when human has to
change the direction of motion and followed by initiation of the head reorientation.
Central nervous system uses two mechanisms to accomplish a turn: foot
placement if planning can be made early and trunk roll motion (piking action
about the hip joint in the frontal plane) otherwise to move COM towards the
new direction.}  of an articulated figure in the single support phase (open
tree structure) in motion is studied in \cite{bmas}. The method consists in
evaluating the influence of the joints
by relating instantaneous joints variations to the corresponding
instantaneous translation of the total center of mass
and a subsequent inversion of the resulting linear transformation in a way
similar to
inverse kinematics.
The authors named the complete process Inverse {\em
Kinetics} since it integrates mass distribution information.
One of the advantages of the proposed approach over inverse kinematics is the
greater convergence rate.
The authors also considered a second order and first order
inverse kinetics formulations.

\subsection {Dynamics Approaches}
 \begin{flushright}
{\small
All principles of gravity are negated by fear.\par
{\em Cartoon Laws of Physics}
}
\end{flushright}

\vspace{.3cm}

The aim of dynamics-based animation is to provide plausible motion
by accounting for the mechanical laws that govern figure's
translation and
rotation. These techniques can be divided into two basic categories:
trajectory-based and controller-based \cite {lasz}. The latter are
discussed in detail in the next section.

Dynamic approaches can be used either for the correction of the existing motion (to generate
more realistic motion by adding constraints) or for directly synthesizing the motion.
These methods are based on applying the laws of mechanics to the articulated skeleton as
an hierarchically ordered set of rigid bodies \cite{Rigbodsim}.
There are different approaches to implement (both direct and inverse) dynamics algorithms.
One of them uses Lagrangian approach and generalized coordinates (each limb angular position
is given in its parent's local coordinate system). In another one \cite{barz} the dynamics
equations for each limb are solved independently and then additional forces that restore the
joints constraints are computed.

Straightforward application of the inverse dynamics methods
to closed kinematics chains is not possible.
The difficulty stems from indeterminancy of the problem: for example, it is
unclear how to distribute force and torque among the legs for an articulated
figure in the double support phase. Ko \& Badler \cite{kb} have suggested to
divide the force in proportion to the distances between the projections of the
center of mass and the ankles. This solution is, however, a static one and
does not account for the dynamics of the motion. Oshita \& Makinouchi
\cite{oshi} have extended this method by including into consideration the
acceleration of the center of mass. Inverse dynamics methods couln not also
handle collisions \cite{park1}.

 One can combine kinematics and dynamics algorithms using the latter
as a post-processing tool that restores the physical realism of the motion by
satisfying corresponding constraints \cite{ko,kb}. It is possible to correct
the results of inverse dynamics computations (if needed), as was done in the
cited papers, to ensure balance and "comfort". The latter is interpreted by
the authors as a limit on the maximum torque that can be exerted at a joint.
If this condition is violated ("strength violation"), the joint trajectory is
modified so as to reduce the torque. This approach is limited, however, since
it is difficult to include some constraints (gravity effects for running or
getting up, for example) \cite{mul}. Moreover, two-stage iterative
computations do not guarantee a convergence.

Another example of coupling kinematics and dynamics computations can be found in
 Newman \& Schaffner \cite {new}.
The authors studied an extravehicular activity via inverse kinematics to
compute the motion of the system (satellite + astronaut) and then used these
recorded motions to calculate the astronaut's body joint torques.
A seven-segmented body model has been used to study movements in microgravity in \cite{zin}.

Isaacs \& Cohen \cite{isa} allowed a subset of the links in the figure to be
controlled
kinematically while the rest were computed using dynamics. A system of
simultaneous dynamic equations has been assembled and the known accelerations
of the kinematics links have been factored out.

In the works of Armstrong et.al. \cite{ams} all the links have been controlled
by the dynamics simulation, however, some have been selected for additional kinematic control
via constraints. Four cases have been considered:

\begin{enumerate}
\item{free link  (no kinematic control)}
\item{fixed in space}
\item{fixed inrelation to the parent joint}
\item{forced to match the position and orientation of the kinematic motion}
\end{enumerate}

A similar approach has been suggested by Westenhofer \cite{west}. A notion of
a {\em kinematic clone} is introduced as a link-for-link copy of the
articulated figure that will be simulated dynamically. Connection between the
figures is provided by the springs joining the corresponding links. Flexibility
of the method could be achieved by  tuning the springs' stiffness.

One of the critical issues in inverse dynamics simulation is the choice of the constraints or
the goal function for the optimization problem.
Li et.al. \cite{li99} have compared different
             optimization criteria in inverse dynamic optimization to predict
             antagonistic muscle forces and joint reaction forces during
             isokinetic flexion/extension and isometric extension exercises of
             the knee. Both quadriceps and hamstrings muscle groups have been
             included in this study. Four linear, nonlinear, and physiological
             optimization criteria
             have been considered. The  authors have not reported
             significant differences in computed joint reaction forces
             and have suggested that  the kinematic information
             involved in the inverse dynamic optimization is more important
             in prediction of the recruitment of the antagonistic muscles
             than the selection of a particular optimization criterion.


  \subsubsection {Spacetime constraints}
 \begin{flushright}
{\small
It is a mistake to look too far ahead.\par
{\em Winston Churchill}
}
\end{flushright}

\vspace{.3cm}

 All the constraint-based approaches considered so far apply constraints to the individual
 instants in time to either compute the needed configurations to meet specified
 constraints (e.g. inverse kinematics), or the required forces to apply at the current
 instant to satisfy the constraints sometimes in the future (inverse dynamics).
 "Spacetime constraints" method introduced by Witkin \& Kass \cite{WK} trying
to compute the figure motion  and time varying forces
 for the whole animation sequence instead of doing it sequentially frame by frame.
 To find the optimal motions, the constraints over the entire motion must be considered simultaneously.
 A placement of a constraint at the end of a motion can affect the behavior of a character at the beginning.
 At the same SIGGRAPH Conference a practically identical method called "Motion
 interpolation by optimal control" has been proposed by  Brotman and  Netravali\footnote{it seems,
  however, that the first variant of the method is cited by far more
frequently}.

 While producing excellent results, spacetime methods are limited to the creation of
 relatively simple motions for the simplecharacters due to high  computational cost
$O(n^2m^2)$, where m is the number of the time steps.

Cohen \cite{coh} suggested to improve the situation by solving the optimization problem sequentially
in "spacetime windows" defined as a sub-sets of degrees of freedom and sub-intervals of time.

A character simplification can be performed using  three basic principles \cite{pop}:
\begin{description}
\item [DOF removal] Some body parts are fused together by removing the DOFs linking them
\item [Node subtree removal] In some cases the entire subtree of the character hierarchy could
be replaced with a single object (usually a mass point with three DOFs)
\item[Symmetry movement] Broad-jumps, for example,  allows one to unite legs into one
\end{description}
Unfortunately, this procedure is hardly could be automated and made motion-independent.

A number of 2D applications have been computed in \cite{ngo} using
 a genetic algorithm\footnote{see details
 in the section {\bf Optimization Methods}} for the solution search.

A cardinal reduction of the computational complexity of the spacetime constraints methods can
be achieved using the hierarchical representation of the optimization problem \cite{lgc}.
Another approach based on fast recursive dynamics formulation allowed the
authors of
\cite{rose} to simulate the motion of the human figure with 44 degrees of freedom.
However, the known drawbacks of these methods (such as an approximate character
of the physical laws satisfaction if the convergence is not attained, for
example) remain.


\subsection {Motion Controllers}

 \begin{flushright}
{\small
 Don't force it; get a large hammer.\par
{\em Murphy's Law}
}
\end{flushright}
\vspace{.3cm}

It should be noted that the word "control" in the computer animation
literature as well as in the present review is used in two different meanings:
\begin{enumerate}
\item {an animator's ability to monitor and modify motion to meet one's
requirements ("Animation is simulation plus control" -- {\em John Platt})
}
\item {a process of maintaining  balance and stability of the animated character's motion}
\end{enumerate}
The current meaning is usually quite clear from the context.

 \vspace{.3cm}

 Advantages of the trajectory-based approaches are obvious: close resemblance to keyframing
(the animator can control the end result), ability to find the most physically plausible
motion even when no accurate solution exists.
The drawback is that a new trajectory has to be generated for each new
instance of the motion. The use of the motion controllers gives a number of
advantages. Controller can be reusable, i.e. allowing to generate a number of
motions with different initial states. There is a possibility to switch
between the controllers during simulation or concatenate controllers to
produce composite motions \cite{ausl}. One of the greatest attractive features
of this approach is an automatic synthesis of motion controllers. And last but
not the least, controllers may be designed in such a way as to mimic some
features of the human neuromuskulo system \cite{bbm}
thus providing a hope for generation of motions that will be not anatomically only but
physiologically (and perhaps psychologically) realistic as well, avoiding excessive
torques, extraneous motions, rapid accelerations etc.

However, the problem of finding a control algorithm (i.e. to answer the
question how to get from EVERY state of the system to EVERY other state)
is more complex than of finding a particular trajectory that reaches a
particular goal state from a particular start state.
As a result, majority of studies of the optimal controllers have been focusing on
the relatively simple systems.

\vspace{.3cm}

A number of different approaches to control the motion have been proposed.
One of the simplest ones is based on the use of finite state machines \cite
{hod}. An algorithm determines what each joint should be doing at every phase
of motion and ensures that the joints perform
appropriate functions at appropriate times. Running, for example,
is a cyclic
activity that alternates between a stance phase, when one leg is
providing
support, and a flight phase, when neither foot is on the ground.
During the
stance phase, the ankle, knee and hip of the leg that is in contact
with the
ground must provide support and balance. When that leg is in the air,
however,
the hip has a different function--that of swinging the limb forward in
preparation for the next touchdown. The state machine selects among the
various roles of the hip and chooses the right action for the current phase of
the running motion.

Associated with each phase are the control laws that compute the
desired angles for each of the joints of the simulated human body.
The
control laws are the equations that represent how each body part should
move to
accomplish its intended function in each phase of the motion.
To move the
joints into the desired positions, the control system computes
the appropriate
torques with equations that act like springs, pulling the joints toward the
desired angles. In essence, the equations are virtual muscles that move the
various body parts into the right positions.

Closed loop controllers were developed by van de Panne et.al. \cite{van1}
for the jumping "Luxo" lamp
and other simple systems using dynamic programming.

For human figures with large number of degrees of freedom the search state is
large. An approach using the fictious external forces called by the authors a
"hand of God" (eliminated, if possible, or reduced
at the late
stages of the optimization) has been used in \cite{van2} to maintain the attitude
of the walking human figure by shifting optimization procedure towards the desired
solution.
The problem with this approach is the incremental removal of the
guiding forces.

A concept of the limiting cycle control to stabilize open-loop trajectories of
the working human model with 19 DOFs has been used in \cite{lasz, lasz1}.
The fundamental entity used in this analysis is a "pose control graph" that is
just a special type of the finite state machine. Each state in the pose control
graph specifies a set of the desired joint angles for the figure with respect
to some fixed reference position \cite{lasz}. Joint angles are driven to their
desired angles by joint actuator torques
generated according to the proportional-derivative law.
The main idea of the approach is to start with a passively unstable system,
divide it into a number of cycles and stabilize each cycle in turn.

A new method for coordination of complex human motion that can be considered as a generalization
of the finite state machine approach has been proposed by Multon et.al. \cite{mu01}.
Each motor unit is considered as an autonomous entity that includes two modules:
\begin{enumerate}
\item{a black box that contains a model and its controller}
\item{a logic representation of the state of the system}
\end{enumerate}

Coordination model is subdivided into three levels with two-way information flow:
 \begin{enumerate}
\item{{\bf the high level:} decomposes the task into the elementary actions}
\item{{\bf the coordination level:} selects the motor units and provides
synchronization} \item{{\bf the motor level:} executes the elementary actions}
\end{enumerate}

As an example the authors have simulated juggling with up to four balls.

A special kind of locomotion controllers can be generated by the evolutionary
algorithms (see \cite{sande} and references therein) that are not directly
applicable to a fixed figure locomotion since they are best for simultaneous
evolution of the character locomotion skill and its morphology.
However, this approach could produce non-trivial results: Hase \& Yamazaki
\cite{hase} have studied biped walking of  humans using genetic algorithms
allowing the shape of the body to adapt for the minimum energy consumption,
muscular fatigue and skeletal load. A model consisting of 10 two-dimensional
rigid links with 26 muscles and 18 neural oscillators evolved from a
chimpanzee-shaped body to a human-like one. A variant of implementation of the
evolutionary algorithm in an interactive frame has been described by Lim \&
Thalmann \cite{lim}.

\vspace{.3cm}

As it has been shown by Playter \cite {pla}, passive dynamics can play a significant
role in the motion stabilization reducing the need for the active control
(a brief summary of the specific movements considered in the cited paper
is given below in the section {\bf {\em Aerial movements}}).
It is also argued that delegating some responsibility for the motion
to the natural mechanical behavior of the body produces a natural-looking,
coordinated movement in contrast to the non-linear control design
approaches that inevitably rely upon the cancellation of some dynamics
effects in order to get a solution and frequently result in a forced
uncoordinated movement. Unfortunately,  the author states,
we do not know for sure if humans are using passive dynamics.
A difficulty stems from the fact that both active control approach and
open loop passive dynamic one could provide a similar dynamic response.
For example, in aerial movements considered in \cite {pla},
both techniques are based on using the  arm tilt to control body twist.

However,  Breni\'ere et.al.  \cite {bre89} speculate
that the differences between the gait initiation by children with under 200 days
of walking experience and adults
( in  contrast to adults, who shift their centre of pressure toward the heels in
           anticipation of the start of the first step, infants do not use
           this strategy) are explained by the fact that children do not
exploit the natural            dynamics as adults do since they have not
mastered            their postural control yet.

Another argument in favor of the passive dynamics can be found in \cite{Eng}.
The authors show that in normal walking use of passive dynamics to control the swing
             trajectory is a mechanism that serves to minimize energy cost
             during locomotion, in addition to reducing the complexity of the
             neural control.
 In a reactive situation (e.g. a slip during
             walking), the energy cost may not be a major determinant of the
             locomotor activity as there is a need for quick corrective action
             under the threat of a fall.  An unexpected mechanical
             perturbation was applied to the foot
             during walking. Video data were input into inverse dynamics
             routine to obtain the joint moment and mechanical power profiles
             and to partition the joint moments into the active and passive
             components. The nervous system still utilized the passive
             dynamics of the effector system: the active control of the hip and knee joints
             were increased but the magnitude of the hip extensor/knee flexor
             moment was invariant and equal to 1.6. The intralimb dynamics
             identified during these responses may serve to simplify the
             complexity of the active control by the nervous system.

The obvious advantage of the passive dynamic control is that it
is not computationally extensive.

Pratt \& Pratt \cite{pratt} have exploited the natural dynamics for the control of a planar bipedal robot.
The authors have exploited the passive swing leg and the compliant ankle limit to avoid the active control.
Simulating 7 links robot they use a finite state machine (with 4 states:
Support, Toe off, Swing and Straighten) for each leg and genetic algorithm to
adjust parameters with the efficiency (defined as distance travelled divided
by the total energy) as the goal function. Another feature non related to the
passive dynamics should be mentioned: introducing a kneecap allows using a
very simple controller (a constant torque is applied so that the knee pushes
against the stop) to get rid of unstable behavior for the straight leg. In
extension of this work \cite{pratt1} a radial basis function neural network
has been used to capture nonlinearities not accounted for by the linear
spring-damper model.

\vspace{.3cm}

An alternative to forward dynamics computations based on using optimized motion controllers
is to increase similarity with the human locomotion and consider advanced models of neuromuscular
systems. This approach encounters a number of difficulties from the poor knowledge
of neuromuscular activity in the living being and uncertainty in parameters to the huge computational
cost. Still, the first encouraging steps in this direction are made \cite{vj}.

A multi-phase optimal control technique has been used to simulate a vertical jump \cite{vj1,vj2}.
The muskuloskeletal system has been described by a set of ordinary differential equations
containing the equations of motion and the first-order equations of the excitation-contraction
dynamics of the muscles. The optimal control problem could be reduced to the non-linear programming
problem (that can be solved by the standard sequential quadratic programming
algorithms). The model of the leg consists of three rigid bodies activated by
nine muscle groups.

A number of papers describe the use of artificial neural nets for the
control of locomotion. For example, a comparison between Multi-Layer
Perceptrons (MLP), Radial Basis Functions (RBF) and Self-Organising Motor Maps
has been performed in \cite {mai} for a spring-legged figure. It was shown that
neurocontrollers based on both MLP and RBF are preferable. These controllers provide good generalization
of the
motion to running at untrained speeds and to running over uneven terrain.

An attempt to develope biology-inspired hierarchical control structure is undertaken in
\cite {craw}. Controllers composed of RBF learn the control required at each level of the hierarchy.
A modified supervised learning algorithm is used for lower levels while a reinforcement learning
approach is exploited for high level.

An hierarchical ("decoupled tree-structure") approach to the development of
controllers for the legged robots have been exploited by  Reichler \&
Delcomyn \cite{reic}. The method could be applied to the robots consisting of a
series of branching chains (like legs or arms) connected to a single reference
member that could be fixed or mobile.
Chains themselves consist of a number of segments (limbs) connected via joints.
The authors have also considered penalty-based and alternative methods to
treat contacts.

Young et.al.\cite{young} have argued that the superiority of the human motor
control system
over controllers developed for robots is in simplification of the
motion commands at the expence of the motion accuracy. The authors
have sugessted two command simplification schemes based on the
equilibrium-poit hypothesis for the human motion control and tested them for
handwriting generation example.

\subsection {Animation of Deformable Objects}
\begin{flushright}
{\small
A cat will assume the shape of its container.\par
{\em Cartoon Laws of Physics}
}
\end{flushright}

\vspace{.3cm}

The topic of this section is partly overlaps with that on the interpolation methods.
It is necessary to mention a few  examples of the unacceptable images
generated by a "brute force" interpolation even based on the underlying
skeleton structure ({\em skeleton-subspace deformation}). This algorithm is
present in the commercial software packages under different names such as
skinning, enveloping etc. \cite{lew}. It is a variant of a more general
approach developed by Magnenat-Thalmann et.al \cite {tal88}. The position of a control point on the deforming
surface of an object lies in the subspace defined by the rigid transformations
 of that point by the relevant skeletal coordinates. In such situations as
shoulders and elbows deformations  this subspace is too limited and no
adjustment of the algorithm weights can produce acceptable  results. The most
common defect is a "collapsing joint" and examples of the partial and almost
complete  collapse of the elbow can be found in the cited paper. The authors
introduce a new method ("pose  space deformation") that generalizes and
improves both shape interpolation and  skeleton-driven deformation techniques.

 A number of requirements to an advanced skeleton-based algorithm has been
formulated by Lewis  et.al.  \cite{lew},
 two of which seem to be the most important:
 \begin{itemize}
 \item {The algorithm should handle the general situation rather than treating each anatomical
 peculiarity as a special case}
 \item {The locality of deformation should be controllable in both Cartesian space and pose space
 (skeleton's configuration space)}
 \end{itemize}

Usually some variant of the multi-layered model  considered first by Chadwick
et al. \cite{chad} is used. These models contain a skeleton layer,
intermediate layers which simulate the physical behavior of muscle, bone, fat
tissue, etc., and a skin layer. Since the overall appearance of a human body
is very much influenced by its internal muscle structures, the layered model
is the most promising for realistic human animation.
Sometimes it is stated that once the layered character is constructed, only the
underlying skeleton needs be animated while consistent
shape deformations will be generated automatically \cite{kar,mtt}.
It should be noted, however, that this approach does not allow  correct
accounting for all effects related to the muscles properties (mass, inertia
moments, volume).

Jianhua \&
Thalmann  \cite{jian} describe an  effective multi-layered approach
for constructing and animating realistic human bodies using metaballs that are
employed to simulate the gross behavior of bone, muscle, and fat tissue. The metaballs
are attached to the proximal joints of the skeleton, arranged in an
anatomically-based approximation. The skin surfaces are automatically
constructed using cross-sectional sampling. Details of the procedure that
provides efficient description of the deformations of a human limb by
manipulating the cross-sectional contours may be found in \cite{kar}.
It includes an  heuristic element (one had to decide which
contours should not be deformed) and thus requires a certain level of the
animator's skill and understanding of human anatomy.

An extension of the free-form deformation model was suggested by Moccozet \& Magnenat-Thalmann \cite {moc}.
The authors enhanced this model using scattered data interpolation on Delaunay
tesselation (Dirichlet/Voronoi diagrams). One of the advantages of the
proposed method is a simple control of the local deformations. As an example, a
multi-layer model is described where "Dirichlet free-form deformation" model is
used to simulate the intermediate layer between the skeleton and the skin for
hand animation.

A use of B-splines for animation of physically-based deformable objects (first of all, muscles)
is advocated by Ng-Thow-Hing \& Fiume \cite {ng1}. Applying a spring-mass model to the B-spline solid, the authors obtained
a dynamic model. They also discussed a possibility
to use nested B-splines to simulate a
hierarchy of anatomical structures within the muscles.

A recent paper by Deunne et.al. \cite {deb} presents a method for animating deformable objects that use an automatic
space and time adaptive level of detail technique and large-displacement strain tensor formulation.
Continuous equations are solved by a local explicit finite element method.
 The method could be extended
to  treat deformation of bodies with topological changes, but, as the authors
stress, this generalization is not trivial.

One of  rather elaborate anatomically correct human models is being
developed
in  Computer Graphics Lab, \'Ecole Polytechnique F\'ed\'erale de Lausanne
\cite {bou, ned2, ned1}. The aim of the authors is to attain fidelity not in
skeleton modeling only, but in description of the body deformations as well.
The model has three levels:  the rigid body structure based on the data from a real skeleton,
 the muscle design and deformation based on physics concepts, and the skin
generation. The skeleton model, in turn, has two levels. The first one
consists of a topological tree structure with generic information about the
joints, their degrees of freedom, their position, the limit angles of each
articulation and so on. The second level is the bones  that are attached to
the joints for animation purposes. The global positions of the joints are
defined based on the reconstructed three-dimensional skeleton while the limits
of the joint angles are fixed from the observation of this skeleton
animation\footnote{it should be noted that, strictly speaking, configurational spaces
of joints could not be assumed independent}.
 The muscle representation is also organized in two levels. In the same
way as the skeleton, one structure has been developed to represent the muscle
actions and attachment to the bones and another one to simulate the muscle
shape.  The deformation has been described with
the data coming from the muscle action structure and with the aid of a
mass-spring model applied on the muscle surface. The physical model is
based on the application of forces over all mass points that compose the
mesh, generating new positions for them. Adding all the applied forces,
the authors obtain a resultant force for each particle on the deformable mesh.
Three different kinds of forces (elasticity,
curvature and constraint force) have been considered.
 The current model is composed of 31 joints with 62 degrees of freedom, 73 bones,
 33 muscles (represented by 105 fusiform muscle parts) and 186 action lines.

 Methods used for simulation of the soft issues of the human body have been considered in detail in \cite
 {maur1} (see also a monograph \cite{maur3}). The author discusses uniaxial and multidimensional models
 that account for elastic or viscoelastic properties of the body parts and are either
 phenomenological or based on the detailed anatomical structure.

A biomechanical musculotendon model that reproduce force behavior observed in experiments
has been developed by Ng Thow hing \cite{vng}.
The model is sensitive to muscle and tendon lengths, muscle velocities and muscle activations.
Deformation rules for generalized cylinders that govern their geometry changes in response to the
underlying skeleton motion are defined.

\subsubsection{Skin modeling}

\begin{flushright}
{\small
In those days the Rhinoceros's skin fitted him quite tight. There were no wrinkles in it anywhere.\par
{\em Rudyard Kipling}
}
\end{flushright}
\vspace{.3cm}

Realistic human animation requires adequate description of the skin
deformation,
especially for the close-ups.

In the {\em elastic surface layer model} by Turnwer \& Thalmann \cite{tur}
the skin was simulated as an elastically deformable surface wrapped around an
articulated figure. This surface was not attached to the underlying layers
(such as muscles and fat) but allowed sliding freely. On the other hand, in
\cite{yasu} the skin points are fixed to the underlying bones and move with
them. Evidently these two examples present extreme possibilities of skin
deformation simulation.

A more justifiable approach is to relate the skin deformation to the
contraction of the underlying muscles \cite{talb}.

An impressive demonstration of skin animation is
modeling of wrinkles evolution that accompanies hand movements \cite{kar, wu}.

\subsubsection {Hair Modeling}
\begin{flushright}
{\small
Would you want to do elastic bodies on each blade of grass?\par
{\em Alan Barr}
}
\end{flushright}
\vspace{.3cm}

One of the most difficult issues in simulating believable
human motion is description of the hair dynamics \cite{had}.
More exactly, the dynamics of {\em long} hair (hair-hair, hair-body and
hair-air interaction) are extremely complex.
A number of approaches have been reviewed in the cited paper.
In the explicit hair models each hair strand is considered for shape and
dynamics that makes them extremely computationally intensive.
In \cite{dal,ros} a mass-spring-hinge model to control the position and
orientation of the individual hair strand has been used.
Another approach \cite{anj} used a simplified one-dimensional beam model.
None of these studies treated hair-hair and hair-air interactions.

A novel approach to this challenging problem has been proposed recently by Hadap \& Magnenat-Thalmannin
\cite{had1}. It combines a description of an individual hair as a serial
rigid multibody chain and continuum description of the hair volume.

The treatment of the evolution of the hair volume by fluid mechanics methods
allows the authors  to consider not only hair-hair interactions but hair-air as
well simulating the motion of the hair-air mixture.

\subsection {Motion Planning}
\begin{flushright}
{\small
If you don't care where you are, you ain't lost.\par
{\em Murphy's Law Book Two}
}
\end{flushright}
\vspace{.3cm}

Motion planning studies originate from robotics \cite{lat}. The problem can
be stated in a classical formulation  as finding
a collision-free path for a rigid or articulated object among rigid static obstacles.
Evidently it is relevant for a human motion animation as well with the only
comment: one has also to avoid collisions among the limbs of an articulated
figure.

Sometimes motion planning as a pure geometrical problem of computing a collision-free path
among static obstacles is referred to as  {\em path} planning \cite{lat1}.

\subsubsection {Path Planning (Navigation)}
\begin{flushright}
{\small
 I know a place round the corner here, where you can get a drop of  \par
the finest Scotch  whisky you ever tasted -- put you right in less than no
time. \par  {\em Jerome K. Jerome}
}
\end{flushright}
\vspace{.3cm}

To give a problem  mathematical formulation a concept of figure's
{\em configurational space} \cite{loz} is useful. A figure is represented as a point
while configurational space encodes figure's DOFs. The obstacles in the Cartesian space are mapped
into forbidden regions in the configuration space. Path planning is thus reduced to finding
a point path in the space with blocked regions. Unfortunately, this space has as many
dimensions as the figure has DOFs.

Huge computational resources needed for the implementation of the rigorous
planners forced development of heuristic methods where, for example, a free
space is represented by a collection of simple cells \cite{bro} or with the
aid of potential field \cite{barr}. However, these approaches fail for path
planning for a figure with more than 4 or 5 DOFs: a number of cells becomes too
large or the potential field has local minima.

Probably the most robust planners for articulated figures with a large number
of DOFs are those based on the statistical approach. A randomized planner
described by Barraquand \& Latombe \cite{barr} in addition to the steps
towards the solution allows "random motions" to escape local
minima\footnote{evidently, an approach identical to simulated annealing is
used.}. A method based on random sampling the configurational space and
connecting the samples in the free space by local paths, thus creating {\em a
probabilistic roadmap} (PRM), is proposed by Kavraki et.al. \cite{kav}.

PRM path planners work well if free space does not contain "narrow passages"
\cite{hsu}. Otherwise,
they require a prohibitively large number of samples. Sometimes \cite{hsu2} it
is recommended to accept the path if the penetration distance into the
obstacle is small (in plain words, a hand can go into the wall if not too
deep).

Sch\"oner et.al. \cite{sch1,sch2} have developed a dynamical system for robot planning. A set of the
behavioral variables (such as heading direction and velocity) define a state space. Path planning
is governed by a nonlinear dynamical system for the behavioral variables.
Constraints (target following, obstacle avoidance etc) are modelled as
the forces that define attractors and "repellers" for the system. Individual
constraints contributions are weighted with weights determined by the second
dynamical system ("task level system") that has a characteristic time scale
smaller than that of the "movement level system" \cite{larg}.

\subsubsection {Collision Avoidance}
\begin{flushright}
{\small
Any body passing through solid matter will leave a perforation conforming to
its perimeter.\par {\em Cartoon Laws of Physics}
}
\end{flushright}
\vspace{.3cm}

A naive algorithm for the collision checking  (to consider every {pair}
of faces or edges in the
object representation) will require  $O(n^2)$ operations and clearly unacceptable for
complex articulated figures. More intelligent approaches should use
the properties of
the objects (or their surfaces) and/or some general space search algorithms
that account for the object location such as hashing, for example.
These methods could be classified as {\em feature-based} \cite{mir} and
{\em hierarchical}
\cite {klo}. The former one exploits the spatial coherence in the geometry
model while the latter pre-computes a hierarchy of bounding volumes for each
object. Hierarchies using different primitive forms have been considered. For
example, spheres are good for a broad class of objects \cite{pal}. These
ideas, of course, could be combined in a single method.

An  algorithm for the self-collision detection of discretized polygonal
surfaces has been suggested in \cite{vol}. The efficiency of the method stems
from two essential issues: 1) the surface is assumed to be regular (smooth)
that allows the authors to exclude a large number of collision tests; 2) a
hierarchical representation of the surface is used.

  A study of cooperative motion between the arms has been performed in
\cite{kog}. The authors rely heavily on the neurophysiology results that state
that the arms movements are primarily defined kinematically. Thus Koga et.al
do not consider dynamics and muscle models but use inverse kinematics
approach in the development of the motion planner.

A generalization of the probabilistic roadmap planner to the case of moving objects has been
proposed recently in \cite {hsu3}. The algorithm is based on sampling the figure's free space with
subsequent integration of the equations and can provide satisfaction of  both
kinematic and dynamic constraints.

An efficient algorithm of the collision avoidance should be built on the efficient kinematic
data structure. Halperin et.al. \cite{halp} called this problem {\em dynamic
maintenance of kinematic structures}. Given a set of rigid bodies linked
together by kinematics constraints (an articulated figure)  that moves in 3D
space, the aim is to efficiently maintain a data structure that would allow
one to quickly answer the range queries as the bodies change their position. A
query usually  specifies a region in space and  asks  whether the figure
intersects this region or lies within a certain distance from it. The data
structure is then used to select a small subset of the bodies to which exact
intersection/distance algorithms will be applied. An initial construction of
such a data structure is a relatively simple problem for a given values of the
joints angles compared to the problem of updating this structure when the
joint parameters change. The authors studied approaches that use balance
decomposition of the tree presenting a figure and got some formal estimates of
the complexity of the problem.

\subsection {Autonomous Character Behavior}
 \begin{flushright}
{\small
In real life it takes only one to make a quarrel.\par
{\em Ogden Nash}
}
\end{flushright}
\vspace{.3cm}

One can distinguish two types of autonomous characters: {\em agents} and {\em avatars} \cite{bad2}.
An agent is a virtual human figure representation that is created and controlled by a computer code.
An avatar is a virtual human controlled by a live participant. A Smart Avatar \cite{shi}
should understand what the human tells it to do. It requires the development of
a conceptual representation of actions, objects, and agents which is
simultaneously suitable for execution (simulation) as well as for natural
language expression. A successfull implementation of such architecture  called
Parameterized Action Representation is reported by Badler et.al. \cite{bad1}.

Another important advancement made in the cited paper is the implementation
of a non-linear animation model via simulated parallelism.  A parallel virtual
machine is called Parallel Transition Network (PaT-Net). Network nodes
represent processes and links contain predicates, conditions, rules, or other
functions that cause transitions to other process nodes. Synchronization
across processes or networks is effected through message-passing or global
variable blackboards. The key feature of PaT-Nets is their conditional
structure. Traditional animation tools use linear time-lines on which actions
are placed and ordered. A PaT-Net provides a non-linear animation model, since
movements can be triggered, modified, or stopped by transition to other nodes.
This is important for autonomous behavior since conditional execution enables
reactivity and decision-making capabilities which one would like to find in an
animated character.

\vspace{.3cm}
The common approach to the behavior of the autonomous character is a multi-level one. These
levels may be called differently and defined in different words but the essence of this
structure is invariant. For example, Blumberg \& Galyean \cite{blu} suggested a
three layer hierarchy:
 {\em motivation, task,} and {\em motor} while in \cite {rey} these layers are named as
 {\em action selection, steering
 ,}
 and {\em locomotion}. These "steering behaviors" are
 largely independent on the details the character's means of locomotion.
 Their combinations can be used to achieve the higher level goals.
 One can list such steering behaviors as pursuit, evasion, quarry, obstacle avoidance, wander,
 separation,
 alignment etc. \cite{rey}.

Blumberg \& Galyean \cite{blu} also state that autonomous character still
should be under some external control (called "directability").
They suggest three levels of such control in accordance with three levels
of
behavior. The imperative forms of the directions are defined as:

\begin{enumerate}
\item{Suggest how an action should be performed if the Behavior System is
wishing to perform this action}
\item{Do it, if the Behavior System does not object}
\item{Do it, independent of  the Behavior System}
 \end{enumerate}

A system for control of behavior of agents described in \cite{grani}
uses both a model of continuous behavior and a discrete scheduling mechanism
for changing the behavior in time.

Goldstein et.al. \cite{gold} have used a dynamic system approach to generation low-level behavior
of autonomous agents similar to one exploited by Sch\"oner et.al. \cite{sch1,sch2}. The authors
have derived a set of ordinary differential equations that govern time variation of the heading direction
and speed of the agent. They simulated such behaviors as obstacle avoidance and target tracking
in two dimensions. The appealing feature of the algorithm is the linear dependence of the number
of operations on the number of obstacles.

A knowledge-based system for the agents behavior in the both "simple" virtual
world and in real world has been developed by Funge \cite{fun}. The models
that able to exhibit low-level autonomous behavior (such as an obstacle
avoidance) are called by the author {\em behavior} ones while for high-level
behavior (a pursuit or evasion) he reserves the term {\em cognitive} models.
One of the main difficulties of following this approach is an appropriate
knowledge representation. The core of the designed by the author  autonomous
"mermaids" and "mermen" consists of the reactive system and reasoning engine.

An architecture fro behavioral human locomotion has been developed by Reich \cite{reich97}.
It includes three main parts: state machine, behaviors and locomotion engine. A number of
behaviors has been implemented either via the locomotion engine or the state-machine layer such as
avoidance, ducking, turning, chasing, terrain awareness etc.

   \subsection {Optimization methods}
 \begin{flushright}
{\small
At the rate you are learning, it would take more than two hundred years to
teach it to you.\par {\em Robert Asprin}
}
\end{flushright}
\vspace{.3cm}

The most computationally intensive part of physically-based animation is by all means the
optimization problem for inverse based approaches (and almost every animation tool includes
some kind of the inverse kinematics solver) or learning (training) for the controller-based methods.

A number of methods have been exploiting in animation from the classical  ones
such as, for example, gradient-based optimization or non-linear programming to
relatively new ones originating mainly from artificial intelligence and
artificial life studies (neural nets, fuzzy logic, evolutionary algorithms).

A trust region method (approximation  of the goal function with a quadratic function in the neighborhood
of the current estimate) allowed the authors of \cite{ude} to solve a rather large optimization
problem with about 4000 parameters. This paper is also of interest as an example of using adaptive
B-spline wavelet to represent joints trajectory.

One of the difficulties of the optimization problem in the animation context is the redundancy
of the basic system (in other words, the system is underactuated \cite
{spong}: the
number of degrees of freedom of the articulated figure can be much greater
than the number of the end-effectors the animator is monitoring). Frequently
the optimization process is got trapped in the local minimum. In such cases
optimization methods that include some stochastic element work best.

Stochastic Gradient Ascent (SGA) has been used successfully to search for locally
optimal controllers for character animation in \cite{pan93}.
The authors established an interesting gait by randomly generating weights for
the sensor-actuator network, and then using SGA to optimize parameters
which affect the quality, but not the type, of the gait. Specifically,
SGA perturbs a random parameter of the controller, and keeps the new controller
if the character's motion gets better. A gait improved if it moved forward faster,
or if the character's body had an upright posture.

SGA has a major drawback. It depends heavily on the initial guess given for the controller.
If you start with a poor guess for the controller, you might not find an acceptable solution
(you get stuck in a local maximum). Such situation is often observed  when SGA
is used for spacetime constraints or forward methods. This is typical for other
methods of searching for a local optimum not covered here - they depend on an
initial guess being near the global optimum in order to find it.

Simulated annealing is designed to overcome this problem, by allowing the
parameters to be varied randomly some of the time, thereby escaping local
maxima. This technique is used to search for a globally optimal solution. It
is similar to SGA, but uses a temperature parameter to determine the
probability that a change of a parameter will be allowed to worsen the
optimization. As a result, it will accept some intermediate steps which worsen
the controller, while allowing the controller to get better eventually. It has
been also used in \cite{pan93} to fine-tune controllers, and has been found to
perform competitively with SGA, converging to a solution a bit faster or
slower, depending on the particular problem situation.

Both SGA and simulated annealing have some drawbacks in common. They depend on a control system
model with a fixed set of parameters to optimize - in \cite{ngo} there are 10 stimulus-response pairs,
and in \cite{pan93} the topology of the control network is fixed: only its parameters change.
The choice of these parameters is critical, and will decide whether the problem is soluble at all.
The parameterizations used in these two papers are extremely rich, allowing for a wide variety
of controllers.

\vspace{.3cm}

 Genetic Algorithms (GAs) \cite{gol} are the methods modelled on a natural
process  which provides adaptation: evolution.
 GAs have been used to solve the problems of producing robust controllers and
 globally optimal controllers found in other approaches to physically-based character animation.
 The reason GAs work is that they belong to a class of a global search
technique. The initial  population takes a random sample of the search space,
usually limited to relatively  simple controllers. Mutation and sexual
cross-over ensure that more complex controllers  are explored as well, often
in widely different parts of the search space.  The search is also guided by
human creativity through the choice of fitness  functions to produce more
interesting and acceptable results.

 It is very hard to specify the torques around the joints so that a character
behaves  the way the animator wants. Torque is not a natural way for an
animator to specify  the motion of a figure. The number of degrees of freedom
of an articulated character  grows very rapidly with its complexity, and it
becomes correspondingly  harder to specify its controls. Even with simple two
or three link characters,  GAs have found some motions which were not
anticipated by the authors \cite{ngo}.

With GAs, the search space does not have to be limited. GAs can combine the sensors,
internal nodes and effectors in a control system in arbitrary ways,
to generate controllers of very different forms than might be designed by hand.

A variable fitness function is also useful for the design of robust
controllers. Gritz \cite{gri1,gri2,gri3} uses a function which starts with a
main goal (reach point X at the end of the time allotment) and gradually adds
 additional style points  for completing the motion early,
penalties for falling over or hitting its head, etc. This allows the main goal
to evolve individuals which are then refined by the style restrictions.
Additionally, fitness is averaged over multiple tries with different
end-target points, resulting in controllers which could move any distance and
stop.

   However, GAs often do not converge to useful results without some tweaking.
In \cite{ngo}     the controller for a bipedal figure was evaluated on the
basis of the forward motion     of the center of mass. The population of
controllers converged very quickly to the strategy     of falling forwards,
and was unable to evolve to a more interesting strategy.     When the
evaluation metric was changed to the forward motion of the point between the
figure's feet,     the GA produced more interesting walking/hopping motions.

In this way GAs are similar to other numerical techniques. SGA is designed only to find a local
optimum, and simulated annealing was designed to address this shortcoming.
Because of their large populations and variety of genetic material available
for solution of the problem, GAs are more suited for global optimization than SGA.
But because they are probabilistic, results from two otherwise identical runs can be quite different,
and we are not guaranteed to find a solution, where SGA will at least find a locally optimal solution.
Gritz \cite{gri2,gri3} showed one way to deal with this problem, by starting the population with an easier
fitness function and gradually adding style restrictions. An alternative that does not use GAs (a "hand of God")
has already been mentioned in section {\bf {\em Motion controllers}}.

\section {Motion Editing}
\subsection{Processing of Captured Motion}

In order to use videocaptured motion for
editing/modification/personification one has to process it.
 Deriving human body shape and motion from optical or
magnetic motion-capture data is an inherently
difficult task. The body is very complex and the data
are rarely error-free and often incomplete. The task
becomes even more difficult if one attempts to use much noisier
video-data instead.

There are two main
approaches for human motion analysis: motion analysis involving human body
parts and tracking of motion \cite{aggarwal99human}. The aim of tracking, using
either single or multiple camera, is to establish correspondence of the image
structure between the consecutive frames. Generally, the features to be tracked
can be chosen rather arbitrary \cite{aggarwal81} trading accuracy for
efficiency. Since human motion is essentially the movement of supporting
bones, it is evident that analysis mode not naturally tied to the human figure
structure could not be considered  an element of the physically based
animation system.

One can distinguish two strategies of the human motion analysis that
explicitly exploit human body structure depending on whether a priori
information about an object's shape is used or not. These approaches are referred
to as model-based and non-model based, respectively \cite{aggarwal99human}.
Both include as the main elements feature extraction and feature
correspondence, the difference being manifested itself in the process of
establishing feature correspondence between consecutive frames. Examples of
non-model based motion analysis using different heuristic assumptions have
been briefly reviewed by Aggarwal \& Cai in the sited paper.

However, model based analysis is by all means the most promising approach for
the processing videocaptured motion intended for subsequent physically based
editing. One can use stick figures, 2D contours or volumetric models. Chen \&
Lee \cite{bb6971} used the model with 17 segments and 14 joints (neck, pelvis,
shoulder, elbow, wrist, hip, knee, ankle) that has been fitted to captured 2D
projections. Similar approach was exploited in \cite{bharatkumar} to model the lower
limbs of the human body. Both papers deal mainly with the gait analysis. An
attempt to describe the human motion solely on the basis of the outline of the
human figure was made in \cite{leung}.

In a number of papers different volumetric models have been used to analyze
the  human motion.  For example, in
the model used in \cite{hogg}, \cite{rohr}  14 elliptic cylinders have been
considered to approximate a human figure in analysis of walking. In the model
\cite{rourke} consisting of 24 rigid segments and 25 joints the surface of each
segment was defined as a union of overlapping spheres. This early model
already included restrictions on the joint angles and the detection of
collisions between the non-adjacent segments.

In all mentioned methods the complexity of matching the image to the human
body model is determined by the number of model parameters. Evidently, fitting
of the image features to the model is simpler for the models with few
parameters (such as a stick figure model), however, extracting these
features can be more difficult and needs care \cite{aggarwal99human}.

A novel approach
for translating the human motion from captured image sequences
to computer animations in real time has been proposed in \cite{nom}.
A motion generator analyzes the current
human motion and/or posture from  data obtained by
processing the source video images, and then generates
a set of the joint angles for the target human body model.
The authors claim that compared with conventional motion capture methods,
this approach is more robust, and tolerant to the broader
environmental and postural conditions.

In \cite{plae} a framework for 3D shape and motion
recovery of articulated deformable objects is developed.
The authors propose a formal procedure that incorporates the use of
implicit surfaces into earlier robotics approaches
that were designed to handle articulated structures.
They demonstrate its effectiveness for human body
modeling from video sequences.

A system for tracking  a 3D person motion that is able tolerate full
(temporary) occlusions is described in \cite{wren}. The system uses an extended
Kalman filter for  probabilistic integration of 2D inputs from two or more
cameras into a dynamic 3D skeletal model. Another approach to motion tracking
that allows occlusion of body parts is described in \cite{yaco}.
the method is based on learning the temporal-flow models using principal
component analysis.

An anatomically based converter for human motion capture has been described
by Molet et.al. \cite{mol}. The method is based on the calibration of both
hierarchical model of the human skeleton and sensors. One of the important
element of the procedure is the transformation matrix between the sensor frame
and virtual joint frame.

\vspace{.3cm}

A closely related question of storing the  produced motion has been studied in
\cite{prod}. The authors compare a number of production/playback methods:
\begin{enumerate}
\item{generating and rendering the motion of characters completely off-line}
\item{the same as the previous, but storing 3D snapshots}
\item {storing the full motion in posture graphs}
\end{enumerate}

The first method (storing 2D snapshot) is the fastest but extremely inflexible.
If the motion is to be used in 3D scene, one has to store different 2D
snapshots for different view angles (for example, in DOOM eight views have
been recorded \cite{doom}).
This last drawback is removed in the second method that still remains a rather
inflexible approach.

The authors advocate the last one since it stores not the {\em results} of
the motions (images) but the {\em motions} themselves (joint angles between
articulated figures). Thus the motion can be easily modified. Another
advantage of this approach is the seamless incorporation of the possibility
to generate motions of simpler articulated figures. This process called by
the authors "slaving" allows one to use for rendering higher resolution models
when
they are close to the viewpoint and lower resolution ones when far away.
The authors used three such models with different level-of-detail (LOD):
"human-high" (73 joints, 134 DOFs), "human-med" (17 joints, 50 DOFs),
and "human-low" (11 joints, 21 DOFs).

\subsection{Editing Methods}
 \begin{flushright}
{\small
Take care to get what you like or you will be forced to like what you get.\par
{\em George Bernard Shaw}
}
\end{flushright}

Motion editing can be performed in several ways.
The simplest one is probably parameterization that hopefully will
allow one to retarget the motion to another character and/or environment.
However, it should be done with care so as not to violate the
motion constraints and not to get such unpleasant artifacts as ground
penetration or feet sliding on the surface.
The problems arise when the captured (or previously generated)
motions have to be adjusted to environments different from
those in which they have been obtained. Rough terrain and
collisions are the two
examples of such potentially dangerous transformations.

Perhaps one of the most trivial editing methods is "motion synthesis by
example" \cite{lam}. The technique is based on choosing the "best fitting"
sample from the set of representative example motions and creating the new
animations by cutting-and-pasting together automatically tailored fragments.

\vspace{.3cm}

One of the the most popular methods for motion editing (more exactly, motion
blending or mixing) is interpolation. Interpolation is used widely for
generating inbetweens during keyframing and for transformation of the graphical objects
(morphing). Usually morphing is implemented as image-based technique \cite{wol}.
A more advanced object-space morphing is based on explicit representation of
the objects (as a rule, either polygons or polyhedra) \cite{gre}. One of the
major phases of the algorithm is establishing of a correspondence between
the geometric features ({\em vertex correspondence problem}). A new
interpolation method that blends the interiors of the geometric objects rather
than their boundaries has been proposed recently by Alexa et.al. \cite{ale}.
As the name of this method ("as-rigid-as-possible shape interpolation")
indicates the aim of the authors is to design a method that would introduce no
excessive global or local deformations.

Motion blending can be performed not in the Cartesian space and time only but
in the frequency domain as well. In the latter case the trajectories undergo
one of the signal processing procedures (it can be, for example, Fourier
transform \cite{unu} or multiresolution filtering \cite{bru95,gene}, ). One
variant of this approach is called "motion warping" \cite{WP}.

Comparing Fourier analysis and multiresolution filtering, the latter is doubtlessly more attractive
for the non-periodic signal of the finite duration. On the other hand, for
cyclic motions such as walking, which are quasi-periodic, Fourier transform
could be an adequate analysis tool.

    The main shortcoming of the geometric interpolation  comes
from independent processing of key figure points that can results in
such artifacts as
extremities shortening.

\vspace{.3cm}

A better way is to interpolate not a set of figure points but the joints
trajectories. However, since it is done independently,  errors\footnote{in addition to the
"static" artifacts such as "collapsing joint" discussed early in the section
{\bf {\em Animation of the deformable objects}} } can be generated:
there is a strong coupling between joints trajectories.

Still, such methods are used up to the present time.  Will \& Hahn \cite{wil} developed an interpolated
synthesis of motion as a tool for producing a continuum space of possible
motions from a small set of examples (that can come from keyframing, motion
capture or physically-based simulations). The authors suggested a hybrid
position and orientation of the skeleton. This hybrid representation does not
ensure constant limb length, but the authors argue that this drawback can be
cured by a post-processing step of conversion to fixed limb lengths. To reduce
the search time for scattered raw data it is proposed to perform a
pre-processing step of filling a regular grid of the parameter space. Linear
interpolation is done for the vector position components (or cubic spline
interpolation in the cylindrical coordinate system) while
spherical linear interpolation is used for each quaternion orientation
component. A number of examples presented in \cite{wil} illustrate the
robustness of the algorithm.
However, as the authors note, the interpolation synthesis is limited to small
number of parameters, on the order of ten (in fact all the examples
demonstrated involved from one to three parameters).

A method called "shape and animation by example" is described by Sloan et.al.
\cite{slo}. Its distinctive feature is the possibility to extrapolate between
the multiple forms and motions. One of the precedents of this method is an
algorithm for multiple blending based on the implicit function presentation of
shapes \cite{turk}. The method of \cite{slo}  can generate a continuum range
of forms ({\em shapes}) and motions ({\em verbs}). The examples to be used for
blending should have the same structure and the same number of degrees of
freedom. Since all forms in shape must have the same topological structure (the
same number of nodes with the same connectivity), the problem of
correspondence between the entities is avoided.
A combination of the radial basis functions and low order polynomials is used
to  perform interpolation of scattered data.
The efficiency of algorithm stems from the fact that one set of linear basis
and one set of radial basis per {\em example} are needed and usually the
number of examples is considerably less than the number of DOFs.
Sometimes the results of interpolation, however, are not acceptable
(such as a muscle bulging or skin shrinking near a the rotation point)
 and the
authors  suggested the algorithms for manual tuning  the process (in
particular, by introducing additional {\em pseudo-examples}).

Recently Brand \& Hertzmann \cite{brand} proposed an approach to the motion
editing based exclusively on the learning
problem.
There is no kinematic or dynamic problem, no underlying figure structure.
All the information is extracted from the processed example motions and
the basic motion ("structure") and secondary motions ("style") are obtained.
The author use a probabilistic finite-state machine (hidden Markov model)
called {\em style machine}.  Some cases of human locomotion are presented.
It should be noted, however, that this apperoach is limited and hardly can
perform
well when transition of the human cyclic motion to the rest occurs:
neither the geometrical nor inertial properties of the human figure
that should be crucial in such transition motion could be learned by
presented algorithm. An evident extension of the method is incorpoartionof the
incremental learning algorithms.

\vspace{.3cm}

Articulated  figure motion editing can be accomplished by the combined use of
direct and inverse kinematic control \cite{bt} already mentioned in the section
on kinematics approaches.
Applying this procedure to an existing motion allows the animator to specify
goal-oriented modifications. The main idea is to consider a desired
joint space motion as a reference model used in a secondary task of the inverse
kinematics control scheme.
This approach provides generalization of the joint-space-based motions to a
larger set of articulated figires.
The process of editing can be also formulated in terms of the COACH-TRAINEE
paradigm \cite{b90} designed primarily as a tool for adding physical realism
to predefined motions via additional constraints: the aim is to retain most of
the kinematics of
the initial motion (COACH) in the process of satisfying a number of
geometrical
constraints to produce a new motion (TRAINEE).

\vspace{.3cm}

Recently Gleicher \cite{gle} has made a comparison of
a number of the constraint-based motion editing methods.
These methods explicitly represent certain features of the motion that are to
be preserved or changed in a controllable way.
In this aspect  the constraint-based  methods differ from other methods such as
pure signal processing approaches \cite {unu, bru95, WP}.

The most effective treatment of the temporal constraints is based on the
observation that high frequencies in motion are almost always noticeable and
thus must be reserved during editing \cite{WP, gle98}.

It should be noted that the constraint-based motion editing methods could not
be applied for editing {\em per se} but as a correction procedure (for
example, to restore constraints after they have been violated by the signal
processing operation).

Gleicher suggested the following taxonomy for these methods based on their
approach to handling temporal constraints:
\begin{itemize}
\item{{\bf Per-Key Methods.} In this case the temporal continuity between the
poses is provided by interpolation.  The author calls  such methods
{\em per-key inverse kinematics}. The main
difficulty of this approach is to get a set of sparse well-chosen key frames to
represent the motion.}
\item{{\bf Motion Warping or Displacement Mapping.} In these methods
\cite{bru95,WP} the changes to the motion are keyframed. This approach is
called {\em motion warping plus inverse kinematics}.}
\item{{\bf Per-Frame Methods.} These methods can be considered as a special
case of the per-key methods where the keys are regularly spaced.}
\item{{\bf Per-Frame Plus Filtering.}
The author argues that the last method is the best among the considered ones.
The first example of this approach has been reported by Lee \& Shin
\cite{lee99} as the "Hierarchical motion editing technique" that used
B-spline fitting to implement low-pass filtering.}
\item{{\bf Spacetime Methods.} In contrast to all the previous methods frames
are not considered individually in this approach. The solver computes an
entire motion (or a sub-window of it)
Unfortunately, a constrained optimization problem turns out to be very large.}
\item{{\bf Physical Approaches.} In the methods considered so far most of
physical properties have been neglected (including Newton's laws) for the sake
of performance: kinetic constraints and energy properties are particularly
computationally expensive in the spacetime framework (that is easy to
understand since such constraints produce {\em coupling} between frames).
However, in some cases it is impossible to discard the laws of mechanics if
one wants to get a realistic motion \cite{PW}.}
\end{itemize}

Modifying human motion capture data using dynamic simulation has been studied by
Zordan \& Hodgins \cite{zor}.  A number of new features added by the authors: a (limited) collision
handler, specialized task controllers to edit a character motion at the
behavior level, and task controllers that allow animation of DOFs that had
not been captured (for example, animation of the whole body with only upper
part captured using a special controller on lower part).
Captured data are converted to the joint angles and used as the desired values for
the tracking controller. The low-level control torques at each joint are
computed using proportional-derivative controller that acts as a muscle model.

In \cite{pol01} dynamics-based editing of the captured data is considered as dynamic
filtering that can be used to maintain physical constraints. A novel element in this paper
is an extension of the technique to treat friction-based contact of the character with environment.

An advanced editing method based on the spacetime constraints proposed in \cite{kom}.
The most appealing feature of this approach is accounting for the muscle properties.
The usual three-component Hill's model is used along with the model that describes the fatigue
of muscles.
The key feature of the computational procedure is a quantitative analysis of
the  (perhaps physiologically implausible) motion:

\begin{itemize}
\item{the given trajectories are approximated by $C^2$ continuous functions (for example,
by B-splines) and differentiated to get velocities and accelerations}
\item{the joint torques are computed via inverse dynamics}
\item{the joints torques are decomposed to the muskulotendon force}
\item{check is performed for every muscle for the entire motion whether the muscle force is less than
the maximum force that is exertable by the muscle}
\end{itemize}
The motion transformation is performed in such a way as to reduce to zero the "infeasibility"
function that measures the extent of the limiting force violation.

 Oshita \& Makinouchi \cite{oshi} have proposed a dynamics-based method for tracking
 and modification of the captured motion. A kinematic motion sequence is used
as an input.  In contrast to the majority of other authors, Oshita \&
Makinouchi do not consider   dynamic simulation as an aid to "cure"
the unrealistic kinematic motion. They rather assume   that the input motion is
already realistic and consider dynamic filtering as means to adjust   the
motion to the changing environment. Primary (human strength model) and
secondary constraints   (accelerations of the center of mass and of
end-effectors) are incorporated.   As an example, an effect of the
load in the hand on the motion of 39-DOFs figure is simulated.

Grassia \cite{grass} has developed a knowledge-based approach to the motion transformations.
The knowledge takes the form od descriptive annotations on the example motions and rules that define
how to combine/modify these motions.The suggested approach contains three main elements:
\begin{enumerate}
\item{motion transformation methods (space warping, time warping etc.}
\item{motion models (domain-specific rules, invariants, high-level parameters)}
\item{algorithms for combining motions}
\end{enumerate}

Lee \cite{lee00} has developed some tools fro motion editing based on the hierarchical approach and
multilevel resolution. In particularity, general method of construction of spatial filters for smoothing
orientation data and hierarchical curve fitting are considered.
A multiresolution analysis of motion used too decompose the motion into the base one and stylistic features.

 \subsection {Personification}
  \begin{flushright}
{\small
There is another: why may not that be the skull of a lawyer?\par
{\em William Shakespeare}
}
\end{flushright}
 \vspace{.3cm}

The amazing thing about human motion is that almost every human has its own
unique style of, for example, walking that can be recognized from a large
distance. The same (perhaps to a lesser extent) is true for
static postures.

One of the essential aims of animation is to provide the character with
individuality.
Leaving aside such attributes as face, hair and clothing,
one has a few ways for personification of the character and its motion:

\begin{itemize}
\item{static biological structure (limbs lengths, muscle parameters)}
\item{dynamic motion individuality directly connected to the
muskuloskeletal system}
\item{additional anthropometric and physiological parameters that can
differ for the characters with the same external (static) appearance:
"comfortable" limiting values of the joints angles, dynamic
muscle parameters, parameters or functions that can account for the emotional
state of the character or the level of fatigue etc.}
\item{specific "secondary" movements that can be superimposed on the (almost
any) basic motion}
\item {motor skill that can be modified during the character life depending on
her/his experience}
\end{itemize}

An example of the adapting a specific motion (running and bicycling) to a new
character can be found in \cite{hodg97}. The characters with different limb
lengths, masses, and moments of inertia are considered. It is stressed that
simple geometric  scaling is not adequate since, for example, a man and a
child differ not only in height but also in proportion (a child has heavier
torso and shorter arms and legs).

The algorithm proposed by Hodgins \& Pollard has two stages. First,
an approximation to the new control system is obtained by scaling based on the
sizes,
masses, and moments of inertia of the new and old models.
Second, a search process is used to tune the new control system.
Since search space for the problem contains a lot of local minima, the authors
used simulated annealing.

A development of this  approach for scaling of the motion was suggested by
Pollard in the later papers
\cite{pol99, pol00}.
To achieve fast performance, she decided to give up complete physical realism.
The task is defined by specifying the reference motion (joint angles and root
position in time), character description (joint locations, degrees of freedom
and physical parameters) and user-defined adjustments(new character
description, new speed etc.).

The algorithm consists of three steps: fitting of a simple task model to the
reference motion data; scaling of both the reference motion and the simple
machine version of this motion; correction of the scaled reference motion so it
matches the scaled motion of the simple machine. The last step -- mapping the
motion back to the complete character -- is performed using inverse kinematics
at each frame of animation.

The task model for running used by the author consists of a point mass
connected to feet by massless springs.
It seems that the most dangerous part of the algorithm is the need to
differentiate twice (!) the motion curves for the character centre of mass
extracted from the reference motion to obtain accelerations -- it is well known
that numerical differentiation of the noisy data is a highly unstable
process proned to errors.

\vspace{.3cm}

An interesting mechanism to introduce individuality into the locomotion style
has been proposed recently in \cite{tak}. The authors developed a corrective procedure that
converts an unbalance motion into a balance one while preserving as much of the original motion as
possible. The method is based on the monitoring and modification of the trajectory of the
 zero moment point. As  by-products of this approach two possibilities to personify
 motion appear. The balancing style of the moving human can be changed either by the variation of
 the size of foot sole or by tuning the joint's stiffness (and thus forcing a model to use waist
 joint and torso joints differently).

 \vspace{.3cm}
 A "physiological" retargetting can be accomplished in the frame of muskuloskeletal human model
 by Komura et.al. discussed above. An animator gets two possibilities to tune the  physical
 ability of the character.
 \begin{itemize}
 \item{by changing the length and the physiological cross-sectional area (PCSA)-- we should refer to
 this approach  as to {\em static} muscle tuning}
 \item{by changing the intracellular pH level as the measure of fatigue --  we
should refer to  this approach  as to {\em dynamic} muscle tuning}
 \end {itemize}
 One can also, of course, increase the PCSA for the certain muscles as a result of the training or
 to reduce the limiting force due to the injury.

 \section{Analysis and Simulation of Specific Motions}

{\small
\begin{verse}
Among terrestrial animals some fly, others move overground either by walking
or crawling; \par aquatic animals either swim or walk. Quadrupeds and myriapods move
with diagonal gaits.\par The lion and the camel amble.\par
\end{verse}
 \begin{flushright}
{\em Aristotle}
\end{flushright}
\vspace{.3cm}
}

One can name a few classes of human movements such as

\begin{itemize}
\item{Posture changes and balance adjustments}
\item{Reaching (and other arm gestures)}
\item{Grasping (and other hand gestures)}
\item{Locomotion (stepping, walking, running, climbing)}
\item{Looking (and other head gestures)}
\item{Physical force- or torque-induced movements (jumping, falling) etc.}
\end{itemize}

It should be noted that, as a rule,  more energetic motions are simpler for
simulation: the dynamics of model constrain the motion and limit the space that
must be searched for control laws for the natural-looking motion \cite{hod}.
For example, it is easier to control running than reaching, and in turn the gymnast's
vaulting is simpler than running in this context: while the gymnast is airborne
greater part of the maneuver duration (control algorithm can influence the internal
motion of the joints only), the runner is considerable part of time in the contact
with the ground and the computed joint torques directly control human motion as a whole.

The most studied specific human motion is, by obvious reasons,
locomotion (walking and running).
Bipedal locomotion is a topic of interest in computer animation,
robotics and biomechanics.

 \subsection {Terrestrial Locomotion}

Different modes of bipeds and quadrupeds locomotion are analyzed in
\cite{mine}.
It is shown, in particularity, that the popular pendulum and inverted pendulum
models often used to study biped gait are limited from the mechanical point of
view. In an ideal pendulum total conversion between potential and kinetic
energies occur ("energy conversion parameter" is equal to 1") while
in walking humans the value of this parameter is about 0.6).
Energetics of human locomotion is usually discussed in terms of the metabolic
energy required for motion that is determined experimentally via oxygen
consumption\footnote{strictly speaking, one should consider both aerobic and anaerobic energy sources
\cite{cape}} \footnote{A direct mechanical calculation of the metabolic muscle activity has been used by
Minetti \& Alexander \cite{mial}}.
The dependence of the metabolic cost on the speed allows one to determine the
transition from walk to run as a speed at which the latter becomes more
economical\footnote{It is interesting to note that transition speeds between
the gaits are different in acceleration and deceleration: "gait transition
hysteresis" is observed \cite{mine}.}.

\subsubsection{Walking}

{\small
\begin{verse}
1 lb. beefsteak, with \par
1 pt. bitter bear every six hours.\par
1 ten-mile walk every morning. \par

\end{verse}
\begin{flushright}
  {\em Jerome K. Jerome}
\end{flushright}
}
\vspace{.3cm}

The human gait is a complex multijoint movement that requires
the coordination of many muscles. It is usually assumed that
uniarticular muscles generate the propulsive energy while biarticular muscles
 provide the fine-tuned coordination.
 Walking can be called "controlled falling." Every time one takes a step, she/he leans forward and
falls, and is caught by the outstretched foot. After her/his foot touches the ground,
the body's weight is transferred to it and the knee bends to absorb the shock.
All the body's limbs, not only feet and hips, but spine, arms, shoulders, and head as well,
are moving to maintain balance. The decomposition of walking to the simpler
movements of different joints has been described by Maestri
\cite {mae}:
\begin{itemize}
\item {The feet and legs --  they propel the body forward.}
\item {The hips, spine, and shoulders --  the body's center of gravity is at the hips;
all balance starts there, as does the rest of the body's motion.
Hips' movement can be considered  as the two separate, overlapping rotations
transmitted through the spine to the shoulders, which mirror the hips to maintain balance.}
\item {The arms --  unless the character is holding something or gesturing,
its arms hang loose at the sides. When walking, they act like pendulums.}
\item {The head and spine from the side --
the spine absorbs some of the shock transmitted to the hips from the legs, making it flex from
front to back a bit. In a standard walk, the head tries to stay level, with the eyes pointing in
the direction of the walk, but it  oscillates slightly to stay balanced.}
\end{itemize}

It should be stressed that in addition to the forward motion in the sagittal
plane a walking human oscillates in the frontal plane. During a step the
body's centre of mass remains medial to the supporting foot and thus the body
is unstable and is falling sideways. It has been suggested that this motion
is controlled in a ballistic manner by proper setting of the initial (toe-off)
position and velocity of the centre of mass \cite{lyon}.

 The complexity of movement is also related to the fact that  a particular muscle can
accelerate many joints and segments, even the joints it does not
span and segments to which it does not attach. A biarticular
muscle can even act to accelerate one of the joints it spans
opposite to its anatomical classification. For example,
gastrocnemius may act to accelerate the knee into extension
during upright standing \cite{ZajacF93}.

It should be stressed that walking is more complex both for analysis and simulation
than running due to the presence of the double support phase \cite {mul}.
Sometimes this forces researchers to use separate algorithms for different walking phases
when simulating the movement of the lower extremity: one
for single support (an open kinematic
             chain)  and the other for the double support phase
(a closed-loop
             one) \cite{Pan88}.
The leg depending on whether it is  in contact with ground or not is said to be in stance
and swing phases, respectively. Detailed description of the locomotion
subcycles can be found in the  paper by Nilsson et.al. \cite{nil}.
Each extension/flexion step is characterized by its duration and associated lower and upper
angle bound values.
When the speed increases, the duration of the double support phase tends to zero
that corresponds to
transition to running.

\vspace{.3cm}

One of the key features of walking is a firm contact of the support foot with the ground
(unless, of course, the character is walking on banana peels). A method to ensure it using
direct kinematics, as has been already mentioned, consists in changing the root limb of the
skeleton when the support foot changes \cite{bru}. Bruderlin \& Calvert have also applied
similar approach to running simulation \cite{bru1}. Control of foot position in inverse
kinematics algorithms is achieved via corresponding constraint \cite {bt96,bt}.

\vspace{.3cm}

Every human has its own recognizable style of walking. Still, to understand and
simulate biped walking one has to search for general relationships.

Analysis of experimental gait data is a hard problem due to a number
of reasons \cite {cha1}:

\begin{itemize}
\item{high dimensionality -- data can include raw (measured) variables
( such as kinematic, kinetic, electromyographic, metabolic,
anthropometric) as well computed via inverse dynamics  parameters\footnote{see, for example, \cite{alk}.
It should be noted, however, that this problem as an inverse one is ill-posed and especially
difficult to solve when the measured data are noisy. A least squares regularization to cope with
problem of reconstruction of torques at joints etc. has been suggested in \cite{kuo}.
}
(joint angles, velocities, moments etc.)}
\item {temporal dependence -- data gathered during walking at a self-selected
pace have a quasi-periodic time dependence
 that hampers the use of classical
methods for analysis of time series (assumption of
stationarity is not valid)\footnote{It explains, btw, that exact repetition of the walking
cycle results in unnatural "sterile" motion, especially when viewed for an extended period of time.} }
\item{high variability -- gait data exhibit intrasubject, intersubject
and between-sample variabilities}
\item{nonlinear relationships between the gait variables due to intrinsic
non-linear character of
human movement dynamics}
\end{itemize}
The recent review of the approaches to gait data analysis, both well
known and relatively new, can be found in \cite{cha1}.
Multivariate statistical methods (principal component analysis,
factor analysis, multiple correspondence analysis),
fuzzy clustering, fractal dynamics, neural networks and
discreet wavelet transform are considered.
Among the most interesting results obtained, an attempt to clarify
the relationships between the muscle activity and the lower-limb dynamics
should be mentioned \cite {sep}.

An extensive review of the mathematical problems related to locomotion control
can be found in \cite {smo}. In particular, kinematics rules and synergetics of stepping
movements in human and animals (from bipeds to myriapods) are considered
and proposed that control of locomotion should be analyzed as a control of
events in space-time continuum. Another mathematical approach to the human
gait based on the dynamics system theory is confirmed by recent observations
\cite {die} where gait transitions are interpreted as bifurcations between
the attractors.

\vspace{.3cm}

A fruitful approach to get insight into the internal mechanics of
the human gait is a comparative study of walking in greatly different
conditions  and/or of different individuals.

Simulation of gait and gait initiation under reduced gravity has been
reported in \cite {bre}.
 A variation of gravity can cause drastic change in the locomotion pattern, as
was clearly demonstrated by astronauts  "walking" on the Moon.
The crucial role of gravity in human locomotion is not easily understood since
it is present in the intersegmental forces relative to each movement and
postural control. The author has considered a simple model of the
double-inverted pendulum with one segment representing head-arms-trunk (HAT)
and the other segment - the legs (LEGS). It has been assumed
that the control
torque between HAT and LEGS segments is proportional
to  the angular variation between segments with frequency dependent
stiffness \cite {bre1}.

It has been shown that the
natural body frequency (NBF) (which had been studied
early by the same author with the aid of inverse dynamics methods \cite {bre96}),
being practically invariable for humans, is
proportional to the square of the gravity.
Moreover, this parameter accumulates kinetic effects of gravity and body
characteristics in oscillating processes of gait and standing pose.
The remarkable consequence is a universal dependence for humans
of the gait initiation duration (the time needed to initiate and
execute the first step) on NBF, whatever will be the characteristics of the
steady gait. This is not true, however, for very young children: as was
shown in an early paper \cite {bre89}, in contrast to adults, who adjust their
gait initiation to match            the target velocity, infants appear to use
one general gait            initiation "program" which takes more than one
step.

Study of walking of children and adults revealed a parameter that is invariant
with age: amplitude ratio of the centre of mass to the centre of pressure
oscillations \cite {bre2}. It should be noted that this amplitude ratio
depends on NBF and thus will change in reduced gravity.

  Understanding of invariant properties of the human walking is useful
as a guiding principle for scaling of the captured or synthesized motion to
different environment or character.

Another important aspect of biped walking is adaptation (adjustment) to the uneven ground.
It was shown \cite {mcf} that  humans use an active
           knee-flexion strategy (with passive hip flexion) for avoidance of changes in terrain height
           while for accommodating  they use an active hip-flexion strategy (with
           passive knee flexion).

\vspace{.3cm}

A comparison of the gait characteristic of a modern woman and
{\em Austalopithecus afarensis} Lucy (AL 288-1)
\cite {lucy}  has been made in \cite {kra}. The model used rigid segments to
represent the foot,
calf, thigh, and pelvis. Head, arms, and trunk (HAT) have been
simulated as a lumped mass placed at the center of pelvis.
The authors computed the energy that a hominid spent in moving
her legs and trunk when walking. Energy is transferred between
kinetic and potential components and between limbs.
Since the system is neither perfectly balanced nor frictionless,
some energy has been added and converted to the cost of locomotion.
The needed segmental parameters (length, mass and mass moment of inertia)
have been taken from \cite{nasa}
for a modern woman and estimated for Lucy.
It was found, surprisingly, that Lucy spent less energy than the (composite)
 modern woman when locomoting at walking speeds.
However, the preferred transition speed from  walk to run
(determined in accordance with \cite{hre} as such a speed that
results in the maximum
acceptable
angular velocity of the ankle joint) is higher for the modern woman.

\vspace{.3cm}

In \cite{bru} a human walking on flat ground has been animated using forward kinematics (procedural)
approach. The basic locomotion style has been specified by
three gait parameters (step length,
step frequency, velocity).

Perhaps mainly due to ease of measurement, the speed-frequency
relation, as determined from individuals walking over a treadmill,
is the most widely used qualitative measure of the human walking,
considered as a characteristic of the "natural" gait \cite {nil}.
It was found that experimental data are fairly
approximated by the power function with exponent value about 0.6.

However, recent experiments \cite{bert} have shown
that the character of these relationships is drastically depends on
the choice of the parameter that is held fixed during experiments
(speed, step frequency or step length).
The authors stressed that conventional theories
(such as pendulum-based models or non-control "passive-dynamic" ones
\cite {mcg}) predict constant frequency independent on speed.
They have succeeded in qualitative reconstruction of all three experimental
curves numerically using constrained optimization.
As the goal function, the metabolic cost of locomotion (oxygen consumption
per unit distance) as a function of speed and frequency has been used.

One of the important by-products of their study is relative insensitivity of
the optimal locomotion on the exact form of the goal function
dependence on speed and frequency (assuming this function is
convex and has a minimum for positive values of speed and frequency).
They also suggested an explanation and a remedy for the optimization
procedure that can remove a slight deviation from observations
(minimum of the metabolic cost function  is somewhat lower than
a preferred walking speed): the only correction one needs to
introduce is to penalize slowness, i.e.
to add to the goal function a term that "rewards" speed.

\vspace{.3cm}

One of the simplest dynamic model of human walking
has been considered in
\cite{bed}. It is a two link planar inverted
pendulum which is made of a single rigid body and a supporting
 masslee
leg. The model behavior has been studied analytically and the knowledge
obtained used for tuning a 15-segment body model.

Garsia \cite{gars} has performed an extensive study of simple passive-dynamics models introduced by
McGeer \cite{mcg} both in 2D (figures without and with knees) and 3D (with no knees). The main
issues investigated have been gait efficiency and gait stability.

Nichols \cite{nich} has studied an approach to gait generation using periodic functions for
specification of the foot's trajectory that results in a scalable gait with the step length,
height and period as parameters. Proportional and proportional-integral
controllers have been implemented for the control of biped walking robot.

An inverse dynamics study of human walking has been carried out in
\cite{fau}. The authors included into the model contacts and friction forces.
Constraints handling has been implemented via Lagrange multipliers. The
simulated gait of a male was in good agreement with experimental data (the
force exerted on the ground during the support phase has been compared).

A hierarchical motion control mechanism has been developed for interactive
animation of human walk in \cite{chu}. As the control hierarchy descends from
the top, automation reduces while more control is given to the animator.
A method computes foot placing automatically planning them two steps ahead of
the current one. B\'ezier curves are used to represent trajectories of the swing
foot and pelvis.

Sometimes the development of the controller for biped walking is based on some
linearized model. Selection of control
             parameters and nominal trajectory determines the quality of
             control and in typical designs, some or all of the parameters are
             selected intuitively. The result is often not the best. If any
             special goal (such as to walk as fast as possible) is desirable,
             the design may become even harder.
             One possible alternative is to obtain the optimal or
             near-optimal design through parameter search.
             In \cite{cheng} a genetic algorithm
             has been used to obtain the optimal design  to achieve
             different goals, such as being able to walk on an inclined
             surface, walk at a high speed, or walk with a specified step size.

  Prentice et.al. \cite{pren} have used a neural network model for generating electromyographic
           patterns observed in human walking. The network has two components.
           First, a recurrent network that receives tonic input is responsible
           for generating sine and cosine waveforms with the proper frequency.
           These waveforms are then shaped by a second feedforward network to
           give the actual electromyographic patterns for eight  muscles of the lower limb and trunk.
           The results obtained using networks with different number of hidden units (4 and 16)
           are compared.
           A model has been able to produce a
           decomposition of central pattern generators  into a self-sustaining oscillatory component
           and a shaping component.
                        In a later paper \cite{pren1} these authors consider artificial neural network
                        models
as an aid for representation of transformation of the kinematic plan into motor patterns.
They have shown that a single trained network could generate muscle activation patterns
for changes in the walking speed, the foot placement and the foot clearance.

                 Three supervised machine
             learning  techniques for prediction of the activation
             patterns of muscles  for walking assisted by a functional electrical
             stimulation ( a multilayer perceptron
             with Levenberg-Marquardt modification of backpropagation learning
             algorithm;  an adaptive-network-based fuzzy inference system;
              a combination of an entropy minimization type of
             inductive learning technique and a radial basis function
             type of artificial neural network) have been studied by Jonic et.al. \cite{joni}.
              The authors have shown that the last method provides the best generalization for
               the
             activation of the knee flexor muscles.

             An approach based on the central pattern generators concept has been used
             to simulate human walking by Taga \cite{taga1}.
             A model included 8 segments, 20 muscles and 7 pairs of neural oscillators .
             In the subsequent paper \cite{taga2} the author has studied the obstacle
             avoidance that has been achieved by modulating the step length to provide a good foot placement
             and by adjusting the basic gait pattern so the swing leg would have a higher
             trajectory over the obstacle. The author has shown that (manually tuned) relatively
             small changes in the joint torques can lead to successful locomotion.

\subsubsection{Running}

\begin{flushright}
{\small
Now, {\em here}, you see, it takes all the running {\em you} can do, to keep in the same place.\par
If you want to get somewhere else, you must run at least twice as fast as that!\par
  {\em Lewis Carroll}
}
\end{flushright}
\vspace{.3cm}

In \cite{hod1,hod} controllers developed as state machines have been exploited to
simulate running. State machines are used to
enforce a correspondence between the phase of the behavior and the active
control laws,  limbs without required actions in a particular state
are used to reduce disturbances to the system, inverse kinematics is used to
compute the joint angles that would cause a foot or hand to reach a desired
location.
Six states have been considered: flight, heel contact, heel and toe contact,
toe contact, loading and unloading. The last two states  of very short duration
are needed to ensure the support (active) leg position before invoking corresponding
control actions.

                                   The
points of contact between the feet and the ground have been modelled via constraints.
It should be noted that such simple description of the contact may be
insufficient for producing realistic running over terrain with variable
properties. The authors of \cite{Fer} assume that the leg and running surface
can be modelled            as two linear springs in series. They use
simulation results as            well as human data to show that runners
quickly adapt their            effective leg stiffness to suit the terrain
stiffness  during steady state running.

The
rigid-body models of the man and woman are realistic because their mass and
inertia properties are obtained using algorithms for
computing mass and moment of inertia of polygonal object of uniform density
\cite{lien}  and density data measured from cadavers
\cite {dem}.   Each internal joint has a very simple muscle model, a torque source.
The runner has 17 body segments and 30 controlled DOFs.

To simplify the problem of specifying the control laws,
several limbs are often used in a synergistic fashion. For example, the
ankle and knee joints of the simulated runner work together to push off the
ground during the stance phase. Wherever possible, the control laws use the
passive behaviors of the system to achieve a desired effect.
During the stance phase, the
runner's knee acts as a spring, both compressing to store the energy and then
extending to release that energy. For the simulation of running, the control
laws make the athlete's arms swing in opposition to the legs.

A  running model based on a one-legged planar hopper with a self-balancing
 mechanism has been proposed in \cite{kang}. A controller for human running
 has been based on the energy consumption.

Influence of inertia on intersegment moments of the lower extremity joints
during running
have been studied by Krabbe et. al. \cite{krab}. The authors have shown that
this effect could be neglected for the ankle but is important for the hip.

\newpage
\subsection {Aerial Movements}
 \begin{flushright}
{\small
 Yet you turned a back-sommersault in at the door -\par
 Pray, what is the reason of that?    \par
{\em Lewis Carroll}
}
\end{flushright}
\vspace{.3cm}

Gymnasts, divers and free style skiers perform extremely complex aerial maneuvers. An airborne
athlete can not apply any external forces or torques to the body so the trajectory of the center of mass
is determined and angular momentum is fixed from the moment of takeoff to the moment of landing,
neglecting aerodynamic forces. However, there are well known aerial movements that at first
sight violate conservation of angular momentum such as: 1) divers leaving the board with rotation about
a side-to-side (somersault) body axis only, can subsequently produce rotation about their head-to-toe
(twist) body axis \cite {fro} and 2) a cat falling with no net angular momentum can straighten
itself in flight to land on all four paws \cite {kane}.

It turns out that an athlete can actively influence aerial motion by the relative movement of the limbs
and torso during the flight. There are three techniques that provide
such control over the motion \cite {pla}:

\begin{itemize}
\item{change of the moment of inertia about an axis by means of grouping the body in order to change
 the rotation rate about that axis}
\item{reorientation maintaining zero angular momentum by performing a sequence of the limb movements}
\item{reconfiguration of the body in such a way that the principle axes of inertia are reoriented
relative to the inertially fixed angular momentum vector}

\end{itemize}

The stability of aerial motion is of primary importance since even
small errors accumulated during a relatively long flight could result
in a disastrous landing. There are two possible control mechanisms
that can generate compensative movements.
 The first one is an active one that assumes that athlete senses
disturbances and generates ("computes") appropriate responses.
Playter \cite {pla} argues that complexity of this feedback control
 approach would force high demands on the athlete's perceptive and motor
control abilities. As an alternative an open loop passive dynamics
control is suggested. Control efforts (torques at joints) during
an aerial maneuver  are just
replayed from memory and compensatory movements are "computed" by the
physical system as part of its natural behavior

The author has shown
that a few gymnastic movements (the tucked front somersault,
the back layout somersault and the front somersault with one and half twist)
could be simulated with passive stabilization that results from
the natural dynamic interaction of the limbs and the body during the flight.
Stabilization arises from the inherent
tendency of the arms to tilt in response to the twisting motion of the body.
The arm tilt forces the principle axes of the system to move in a direction
that compensates for tilt and twist errors. The author proved the possibility
of passive stabilization
using a simple model: the head, torso and legs are
considered as a single rigid body while arms could be raised and lowered.
Torques at the shoulder joints are provided by torsional springs and dampers.
The model (neglecting translation of the centre of mass) had five
degrees of freedom: three for the rotation of the body and two for the
relative motion of each arm \cite{pla}. This simple three-body model
exhibits stable somersaults if spring used at the shoulder is adjusted
correspondingly. It should be noted that  Yeadon \& Mikulcik \cite{yead} do not accept the generality
of the Playter's conclusions. They agree that passive stabilization can be effective
when the arms are abducted more than $30^o$ from the midline of the body.
It is, however, insufficient when somersaults are performed with the arms adducted close to the body
and then stability could be achieved using proportional-derivative controllers.
The authors have studied different strategies for controlling twist in twisting and nontwisting
sommersaults using a 11-segment model: 1) symmetrical arm abduction; 2)
arching of the body; 3) asymmetrical arm adduction/abduction.

A simulation of the female gymnast performing a vault from the spring board
over the vaulting horse has been reported in \cite{hod}.
The model has 15 segments and 30 DOFs. State machine was used to control
the motion with the following states: hurdle step, board contact, first flight,
horse contact, second flight, landing.
 \vspace{.3cm}

A rather sad example of the human aerial movement is reported in \cite{hatze}.
The author, being called as an expert to the court,  simulated an overrotated rock'n roll
Betterini somersault that ended by the severe injuries of the female dancer when she hit the floor
with her chin.
\vspace{.3cm}

A simulation of a human diver with 38 controlled
degrees of freedom has been described in \cite{hod2}. The human model can
perform a number of 10 meter platform dives. The dynamic model of the diver
consists of 15 rigid bodies connected by the rotary joints. The dynamic
properties of the rigid bodies were calculated as for the runner model
described in the previous section.

The control system for the diver is hierarchical. The low-level control is
provided by proportional- derivative servos that move the joints towards their
desired positions. Balance on the diving board is provided by a controller at the
ankle that computes the angle for the ankle that would place the body's center
of mass over the feet. This angle serves as a desired angle for the low-level
PD control. High-level control for the dive is provided by a state machine
that alters the desired configuration of the diver. Five states are used in
the 10m platform dives: Compression, Decompression, Flight-Phase 1,
Flight-Phase 2, and Entry. The high-level control alters not only the desired
values for the joints but also the gain on the low level PD servos. For
example, the gains required for the compression phase of the dive are higher
than the gains required for the flight phase. The gains and set points for the
controllers were tuned by hand to ensure that the diver performs the dive and
enters the water vertically.
Still the model of muscles used by Wooten and Hodgins was very simple (a torque
source at each joint) and the strength of the individual joints was not taken into account.
This simplification, as the authors themselves noted, could results in a simulated
human that is stronger or faster than a real one and can perform motions impossible for humans.

\subsection{Arm Movements}
\begin{flushright}
{\small
neither shall ye touch it \par
{\em Genesis, 3, 3}
}
\end{flushright}

\vspace{.3cm}

A great variaty of human arm movements can be classified into a number of
groups such as

\begin{itemize}
\item{Goal-directed manipulation:
\begin{itemize}
\item{changing position: lift, move, raise, push, pull, draw, jerk, toss,
throw, hurl, thrust, shake etc.}
\item{changing orientation: turn, spin, rotate, twist}
\item{changing shape: fold, stretch, bend, spread, squeeze etc.}
\item{contact with object: grasp, seize, grab, embrace, grip etc.}
\item{joining objects: tie, nail, sew, bind, pin, envelop etc.}
\end{itemize}}
\item{Empty-hand gestures: wave, snap, point, show, shrug etc.}
\item{Haptic exploration: touch, stroke, knock, throb, tickle, hit, slam, tap,
kick, pat etc.}
\end{itemize}

karla et.al. \cite{kar} cite another taxonomy:
\begin{itemize}
\item{{\bf Semiotic:} to communicate information}
\item{{\bf Ergotic:} to manipulate and create objects}
\item{{\bf Epistemic:} to explore environment}
\end{itemize}

In this section gross arm motion is considered. The discussion
of the
animation of fine hand movements such as grasping, for example,
can be found in \cite{hand1, moc}.

\vspace{.3cm}

Reaching motion is obviously one of the most frequent arm movements.
This motion has been studied extensively due to both the relatively simple
experimental setting and its importance for ergonomics.

The general features of the reaching motion are well known \cite{chaf}:
\begin{itemize}
\item{the trajectory, if unobstructed, is nearly straight.}
\item{the velocity profile is bell shaped.}
\item{the acceleration profile has two peaks.}
\item{any mass added to arm tends to reduce the velocity and acceleration
values}
 \end{itemize}

It is also well known (starting from N. Bernstein observations in 1920s)
that the straight trajectory of the end-effector
does not prevent the trajectories of the intermediate joints to be very
complex\footnote{The problem of understanding how human body overcomes an
excessive number of degrees is called sometimes "Bernstein's problem"
\cite{chaf}}.

Numerous experimental studies of hand periodic ("rhythmic") motions revealed
the power law (referred to as "2/3 Power Law") relationship between
the tangential velocity and the radius of curvature of the end-effector
trajectory\footnote{In fact  2/3 power law is formulated for the angular
velocity $a$ proportional to  the curvature $c$ of the end-effector path $a
\sim c^{2/3}$. The dependence of the tangential  velocity on the radius of
curvature is written as $v \sim r^{1/3}$.}.  It has been considered as a
fundamental invariant of the central nervous system  in the formation of
rhythmic endpoints trajectories and even as an assessment criterion for the
quality of motion models \cite{scha}. However, Schaal \&  Sternad in the just
cited paper published this year have shown that this law  is valid only for
planar trajectories of relatively small size almost exclusively  studied in
the precedings works. Analysis of the 3D endpoint trajectories demonstrates
that deviations increases with the pattern size.

The authors explanation of this disagreement is based on the hypothesis that a power
 law is a by-product of the smoothness of the endpoint trajectory. A similar opinion is
 supported by Todorov \& Jordan \cite{tod} who used in their studies "minimum jerk"
 optimization criterion (which is equivalent to minimization of the higher frequency
 components of the trajectory).  The crucial step in Schaal \&
 Sternad analysis is to assume that smoothness is implemented in the intrinsic
joint space coordinates.  For small patterns it results in smoothness in
Cartesian coordinates due to linear  transformation between the joint
coordinates and endpoints coordinates. The behavior for large  patterns is
explained by the non-linearities of these transformation.  The authors
righteously argue that these results indicate the primary role of the joint
space  in motion generation.  They, however, warn that smoothness in joint
space (and, hence, minimum jerk movements  and minimum torque changes) is a
criterion valid for rhythmic movements that could fail for  discrete
movements\footnote{since rhythmic motion is phylogenetically older than
discrete one,  they could use different neural circuits \cite{scha}.}.

In  another paper \cite {ster} these authors have analysed two hypothesis
of movement segmentation of  endpoints trajectories in human drawing movements.
The kinematic characteristics of the endpoint trajectories and the seven joint
angles of the arm have been studied.
The authors have concluded that while the endpoints exibit segmentation, the
joint angles
show continuous oscillatory patterns.
This phenomena is explained by the nonlinear transformations of the forward
kinematics of the arm. The authors suggest that the rhythmic drawing movements
are
described best in terms of continuous oscillatory pattern generators in joint
space.

It is worth to note that joint space coordinates are thought to be the
coordinates use to encode reaching movement direction ("spatial preferred direction")
 in the primary motor cortex  of primates
\cite{aje}. The authors of the sited paper considered three coordinate systems
(in addition to the joint coordinates, Cartesian  and shoulder-centered)
2-DOFs planar arm.
The joint coordinates best explain  observed cellular response patterns data for
the curved motion
experiments.

The framework of the equilibrium point hypothesis has been used by Latash
et.al.
 \cite{lata}  to reconstruct equilibrium trajectories of the elbow and the
wrist during fast voluntary movements. An explicit relation between control
variables to the wto joints has been suggested as the implementation of the
symple synergy. Sainburg et.al \cite{sain}
have studied reaching movements and have found that a three-stage
system sequentially links anticipatory, error correction and
posture mechanisms to control intersegmental mechanics.

A model for simulation of reaching movements that contains 2-DOFs planar
manimulator, a Hill type model od 6 muscles and artificail neural net to
represent the nervous system has been developed by Karniel \& Inbar
\cite{karn}. The results obtained are considered by the authors that the
cental nervous system able to generate typical reaching movements with a
simple feedforward controller that controls only the timing and amplitude of
rectangular excitation impulses to muscles. It has been also shown that
nonlinearity of the muscle properties is essential for achievement of such
type of control.

Over-arm throws constrained to the sagittal plane have been studied in
\cite{chow} using a muscle-actuated two-segment model representing the
forearm  and hand. The moments at the joints were produced by the muscle-tendon
models representing the net action of the elbow extensors and wrist flexors.
The controller had to specify the times of the activation of each muscle-tendon
and the time of release of the ball. The study had shown that the timing
accuracy of the muscle action should be rather high.
The task was completed in less than 150 ms which is too fast for afferent
signal feedback to be received, processed, appropriate efferent signals sent
to the muscles and a change in the force occurs.
Thus, as the authors argue, such muscle action must be pre-planned.

A simple model of a two-segmented arm with only two muscles
have been used in \cite{luss} to study an underarm throw and overarm stroke.
They authors have found that, due to series tendon compliance, there are often
two distinct optimal delays for a muscle with given moment arms.
The choice of a global  optimum out of two depends on the strength of other
muscles and the weight of a load, usually a shorter delay being advantageous
for heavier load.

Different cyclical elbow-wrist movements, including unidirectional rotations
at the elbow and wrist in the same direction, bidirectional and free wrist
motion, have been studied in \cite{doun}.
The authors have shown that control at the elbow was principally different
from control at the wrist.
Elbow control in all three movements studied was similar to usually observed
during the single-joint movements and largely independent of wrist motion.
The elbow muscles were found to be responsible not only for the elbow motion,
but also for generation of torques that play an important role in wrist
control. These torques are ther primary source of wrist motion. The authors
have  proposed  a hierarchical organization of the elbow-wrist coordination:
the elbow muscles generate movement of the whole linkage while the wrist
muscles provide a fine correction of the motion.

The patterns of joint kinematics and torques of planar arm reaching movement
in the sagittal plane have been considered in \cite{gott}.
Dynamic muscle torques have been calculated via inverse dynamics method
neglecting gravity.
The dynamics components of the muscle torques at the elbow and the shoulder
have been almost linearly related to each other for the movements in almost all
directions.
Both have been similarly shaped symmetrical pulses. This effect is called by
the authors {\em linear synergy}. The relative scaling of the torques at two
joints has been found to depend on the movement direction.
The authors argue that their results, not being incompatible with other
hypotheses of motion coordination, are best explained assuming that
voluntary, multiple degrees of freedom rapid reaching movements may use
rule-based, feedforward control of dynamic joint torque.
This control system may operate in parallel with a positional control system
that maintain body balance.

A detailed model of the elbow, forearm and wrist has been developed in
\cite{lema}. The model contains the humerus, radius, ulna and hand connected by
the appropriate joints. Twenty muscles have been included into the model. They
were simulated by a force producing contractile element in a series with a
tendon
that has been modelled as a spring of constant stiffness.
The unusual feature of the model is the account for the limit joints angles
via passive moments added at the joints. These moments have been determined
using experimental information. One of the conclusions of the authors is the
high
sensitivity of the model to the muscle length and tendon slack length.

The biologically motivated approach to the automatic generation of the lifelike gestures is
proposed by Kopp \& Wachsmuth \cite{kopp}. The authors have combined knowledge-based techniques
 with a trajectory generation method based on parametric curve composition using splines.
 The articulated figure used in the cited paper contained 43 DOFs for the main body and 20 DOFs for each hand.

Kinematics of reaching has been studied by Hestens \cite{hest1}
using developed by him geometric algebra approach (see the section
{\bf Kinematics approaches}). The author follows Bullock and Grossberg
\cite{bull} assuming that, since the experimentally observed hand trajectory is
straight, hand position end points are {\em control variables}  for reaching
movements. A general formulation of the kinematics of reaching is presented
and solution of the inverse kinematics problem that parametrizes the joint
angles in terms of the wrist position is derived. An advantage of the proposed
formulation of the basic {\em reaching equation} is its coordinate-free form.
As the control variables {\em the target direction} and {\em the arm
extension} are used.

The author also considers general questions of quantitative description of the
neural sensory-motor system. Since body kinematics is the geometry of
movement, the accurate motor control could be achieved if this geometry is
reflected somehow at all the levels of sensory-motor  processing. He advocates
the use of the term "neurogeometry"  \cite{pell} to describe the mathematical
theory of motor control.

Schouten et.al. \cite{scho} have studied the effect of the frequency on
the reflexive feedback gain of human arm using a 2-DOFs model with six
muscles. A two-step procedure has been used: 1) optimization of static muscle
activations; 2) optimization of reflex gains using an "additional control
effort" criterion. The results show that for a given posture shoulder muscles
have the largest contribution while the biarticular muscles have a relatively
small contribution to the behavior. The modest deviations from the
experimental data are explained by the unmodelled mechanical effects of
the cross-bridges in the Hill-type muscle models.

Chi et.al. \cite{chi} have animated arm movement using EMOTE (Expressive MOTion
Engine) system developed by the authors recently. The system defines motion
in terms of Laban motion analysis notation in which, for example, "Effort"
has four attributes (Space, Weight, Time, Flow) while "Shape" has
Horizontal, Vertical and Sagittal "dimensions". The arm model has 1-DOF elbow
joint and spherical 3-DOFs shoulder and wrist joints.

\subsection {Miscellaneous Movements}

 Various planar gymnastic exesizes (transversing a set of irregularly-spaced monkeybars,
climbing a ladder, fall recovery have been simulated in \cite{lasz2} using
motion primitives and state machine control. The character of these primitives
for, for example, monkeybars is as follows: "grasp closest rung",   "grasp
rung following closest rung", "pull up using support arm" etc. As the authors
state, this approach is still far from reproducing complex 3D human motions.

The difficulty of simulations of such motions as flips and jumps arises from
the need for an accurate timing for a successful execution. In \cite{hua} a
decision-tree search algorithm has been used to model such motions in 2D.

   The model of the leg consisting of three rigid bodies activated by nine
muscle groups has been    exploited in \cite{vj1,vj2} to study a vertical
one-legged jump. Three phases have been analyzed    starting from the squat
position till the stabilization of the body after landing.

Quasi-elastic behavior of joints in such movements as jumping have been
investigated in \cite{seyf} using an elastic three-segment leg model.
The effective operation of the muscles crossing the knee and ankle joints has
been described in terms of rotational springs. The authors have found, in
particularity, that a three-segment leg tends to become unstable at a certain
amount of bending expressed by a counter-rotation of the joints; nonlinear
joint torque-displacement behavior extends the range of the stable leg bending;
biarticular structures (such as a human gastrocnemius muscle) and geometric
constraints support homogeneous bending in both joints. The future study
should account for the segments masses and moments of inertia and dissipative
joint operation (such as a heel pad deformation).

A movement strategy of the human forward trunk bending has been studied in
\cite{al}. The authors note that two functionally different behaviour goals
are achieved during this movement: the bending itself and the maintenance of
the equilibrium.
They try to decompose the motion into three dynamically independent
components connected with the hip, knee and ankle movements. The results show
that the hip and the ankle movements can be treated as the independently
controlled motion units and it is proposed that central nervous system controls
this motion sequentially to perform the bending and to maintain equilibrium.

A planar model for lifting activity has been used in \cite{ayo}
to check a hypothesis that
the body tries to minimize the work to be done. The goal function has been
also subjected to a number of constraints that include the physical
workplace and maintenance of balance. The results computed for the body model
containing 5
rigid links were in good agreement with experimental data.

A simulation of the finger motion in the saggital plane assuming minimum
of the expenditure of energy is done in \cite{sho}. It it difficult to
understand what model of a tendon was really used: the authors refer to the
tendon
as "an inextensible filament" in the model description, but later speak about
"changing of tendon length".

Hestens  has studied the oculomotor system for accurate gaze
control and in par�icularity
 {\it saccade} \cite{hest1}. Saccade is a rapid shift of gaze in order to
fixate a target object in the visual field. For kinematic purposes, the eye is
simulated as a ball in a socket joint, so it has three degrees of
freedom. There exists a number of experimentally proven laws of
neurobiology. One of them (Listing's law) asserts that there is a
unique gaze direction, called the {\it primary direction}, such
that any saccade is obtained from it by parallel transport along
the geodesic. The main body of the paper is devoted to
investigating implications of Listing's law. The main problem is
to understand what control parameter central nervous system uses to
control pursuit of the moving object with gaze. In the considered
paper the optokinetic model of the human gaze is developed and
analyzed.

Ballroom dancing studied by Lake \& Green \cite{lake} gives an example of a motion
which decomposition into primitive movements is relatively simple and can be done
hierarchically. Each dance can be divided into a set of motion sequences called
{\em patterns} with a typical duration of 2 to 10 seconds. Each pattern, in
turn, consists of a sequence of {\em positions} that define location and
orientation of the body after a specified time interval (step). Almost all
patterns are composed of the transitions between the five fundamental
positions \cite{lake}. Motion control used by the authors is based on a finite
state machine approach. A limb may be in one of three motion states:
\begin{enumerate}
\item{"Free swing": all internal torques are set to zero.}
\item{"Move": torques should provide smooth transition of the limb from the starting to the final
position within the specified time limit.}
\item{"Maintain": as the name suggests the limb should be kept in a fixed angular position relative to its
parent, thus large restorative torques are applied in case of deviation from the current position.}
\end{enumerate}

The task manager calculates the time needed to complete the step and initiate timers for all links whose
state shoud be set to "Move". Rotational matrices describing the current position are converted to
the quaternions since interpolation along the quaternions curves is preferable.
The authors encountered a problem in the implementation of their method for articulating relatively simple
figure that manifests itself as limb oscillations leading to numerical instabilities. The remedy used
in the paper is hardly a good one: moments of inertia for each link were multiplied by a factor of 300 (!).
\subsubsection {Multilegged Locomotion}
{\small
\begin{verse}
A large amount of nonsense has been said about this subject, not only  by \par
ordinary people but even by excellent scientists and anatomists, who prefer \par
to pass on incorrect second-hand theories, rather than trust their own observations.  \par
\end{verse}
 \begin{flushright}
{\em Giovanni Alfonso Borelli}, De Motu Animalium (1680)
\end{flushright}
}
\vspace{.3cm}

Multilegged locomotion is not directly related to the human motion (except in
cases when a human uses all her/his extremities to achieve a greater stability
of gait). Still, the studies of this topic are of considerable interest due to
the differences of problems of, first of all, motion controllers development
and other characteristics, and enlarge our knowledge in biped locomotion as
well.

Different approaches to neuromechanics of legged locomotion have been analyzed by
Full \& Koditschek \cite{fullk}. The authors have listed 12 hypotheses of  the organization
of legged locomotor systems. They suggest the hierarchy organism $\Rightarrow$ anchors
$\Rightarrow$ templates.
"Templates" are the simplest models that exibit a desired behavior and disregard skeletal
type, leg number etc. Examples of templates are a well-known model of spring-loaded
inverted pendulum that describes motion in the sagittal plane and proposed by the authors
lateral leg spring model that simulate animals bouncing from side to side.
"Anchors" are elaborate models grounded on the morphological and physiological data that should
account for the multiple legs, joint torques that actuate them, the recruitment of muscles and control
by the nervous system.
The creation of the templates is aimed at the resolution of the redundancies of multiple legs,
muscles and joints by exploiting synergies and symmetries.

Surprisingly templates turned out to be rather universal for the terrestrial locomotion.
Moreover, such a parameter as a relative leg stiffness in the spring-loaded inverted pendulum model is similar in
six-legged trotters (cockroaches), four-legs trotters (dogs), two-legged runners (humans)
and two-legged hoppers (kangaroos) \cite{fullk}.

Walking of quadrupeds is similar to that of humans and can be considered as
two bipeds walking in front of the other, moving their limbs with a phase
shift of 3/4 of the cycle \cite{mine}. Similarily, a trotting quadrupeds can be
presented as two running bipeds with a phase shift of a half of the cycle (or
zero for gait of quadrupeds called the  rack). Quadrupeds also exibit a few
variants of high-speed gait -- gallop --that is analogous to human skipping
(the
latter is not a favorite humans locomotion gait since it is {50
expensive energetically than running at a comparable speed \cite{mine}).

Statistically stable multi-legged motion has been studied in \cite{mck}.
It has been shown that when more than two legs are always support the
articulated character, the center of mass is generally lies in the sustentation
polygon.
In this case in contrast to the human walking the balance control is a less
important issue than the coordination of the legs.

In \cite{cym} a model of six-legged walking is described optimized by
           reinforcement learning  to achieve stable motion.
           The cost function is  based on the frequency
             of the model's loss of stability evaluated for randomly chosen
             initial leg positions. Genetic algorithm has been used at initial stages of optimization
             to get to the vicinity of the global minimum followed by a Monte Carlo
           method to search a smaller space of solutions.

           It should be noted that the six-legged locomotion is unique in that sense
           that absolutely stable motion is possible (at least theoretically,
           assuming ideal coordination) when alternatively three legs are in the support phase.
           If alternating tripod gait (first and third left legs move in phase with second right one
           all three in anti-phase to the opposite tripod) a tripod acts as a
single virtual leg and            coordination of the motion is similar to that
of bipeds \cite{full}. Six-legged insects tend to have the shortest swing
period possible and as speed increses the swing period changes little and the
stance period decreeses \cite{klav}.

           Mullerwilm \cite{muller} has simulated the coordinated
             interaction between the walking legs of a multi-legged animal using
             a neural network consisting of the separate modules with
oscillatory              capabilities. A variant  of reinforcement
             comparison learning has been used to train the network.

Trajectory-based optimization of the quadruped locomotion has been reported in
\cite{tork}. To simplify the problem the authors restricted themselves to
optimization of the motion trajectories of  two key mass points instead of
all degrees of freedom.
Another remarkable feature of their work is, as they formulate,
"the capability to stretch the laws of physics if necessary": both physics and
"comfort" are treated as constraints that can be satisfied only approximately.

The simplified model consisted of two point masses connected by a spring that
served to model the internal forces of the back of the quadruped.
The trajectories have been represented as $C^2$ piecewise-continuous splines.

Animation of the motion as well as animation of the growth of animals has been
studied in \cite{walt}.

Ito et.al. \cite{itos} have developed  a  central pattern generator model
that can account for periodic perturbations from the environment and adaptation of
quadruped locomotion.
The  model accounts for  three
             types of dynamics: environmental, rhythmic motion, and
             adaptation dynamics. Authors argue  that the time scale
             of adaptation dynamics should be much larger than that of rhythmic
             motion dynamics.

Klavins et.al. \cite{klav} studied two schemes for coordination of
multilegged locomotion (central pattern generator and reflex driven
coordination) using two simple models:
\begin{enumerate}
\item{"bipedal bead on a rail" (BBR)}
\item{"bipedal spring loaded inverted pendulum" (BSLIP)}
\end{enumerate}
One of the aims the authors declare is to develope an "interpolation"
between the pure feedforward and pure feedback coordination mechanisms.

\subsection {Large-scale Numerical Simulation of Human Motion}
\begin{flushright}
{\small
 And what is the good of a small dare, Roger?\par
  {\em William Golding}
}
\end{flushright}
\vspace{.3cm}

Detailed models of muskulotendon system allow researches to perform very
accurate computations of human motion. Unfortunately,  these models
are extremely computer intensive.

A number high-performance computations of human motion were carried out by
Anderson \& Pandy and their colleagues. The model complexity so far ranges
from 14-DOF actuated by 46 muskulotendon units \cite{and1} to 10-segment,
23-DOF skeleton actuated by 54 muskulotendon units \cite{and}.
The head, arms and torso (HAT) in the last model were lumped into a single
rigid body, and this segment articulated with the pelvis via a 3-DOF
ball-and-socket joint. Each hip was modelled as a ball-and-socket joint while
a knee as 	a 1-DOF hinge. Two segments were used to model each foot.

Each leg was actuated by 24 muscles, and the movements of pelvis and HAT were provided
by 6 abdominal and back muscles.
Each actuator in the model was
represented by a three-element Hill-type muscle in series with elastic tendon
\footnote{It has been shown
in \cite{hatze} that Hill-type models are oversimplified and sometimes produce grossly erroneous results.
}. Parameters describing the
mechanical properties of each muscle were obtained from the
literature and scaled to the strength of individual subjects.
The interaction of the feet with the ground was simulated using a set of spring-damper
units distributed under the sole of each foot.

To solve the optimization problem the authors of \cite{and} discretized the excitation history
of each muscle
into a set of 15 control nodes which became independent variables.
A minimum
of metabolic cost per unit distance had been chosen   as a goal function.

\vspace{.3cm}

It is useful to cite some figures that show the scale of computations.
A 14-DOF model required about 3 months of dedicated processing time on SGI
Iris 4D25 or 77 and 88 hours on Intel iPSC/860 and Cray Y-MP 8/864,
respectively \cite{and1}.

The authors have analyzed performance of different computers using a conventional serial machine,
a parallel-vector-processing machine, and a
multiple-instruction/multiple-data (MIMD) parallel machine.
The serial optimal control algorithm decomposes very efficiently into a parallel algorithm.
It
consists of three parts: (1) forward simulation, (2) computation of derivatives, and
(3) parameter optimization \cite{and2}. Of
these three parts, computation of the derivatives dominates total CPU time (i.e., over 90 percent).
When
implemented on a MIMD supercomputer, computation of the derivatives scales almost linearly with the number of
processors used. Specifically, on any MIMD computer, the  authors were able to
compute derivatives 100 times faster with 128
processors than with just 1 processor.
In contrast, the Cray performed best during parameter optimization of the
controls, executing the parameter optimization routine 37 times faster than the serial machine.
The ideal computer architecture for solving very-large-scale
optimal control problems appears to be a hybrid system in which a vector-processing machine
is integrated into the communication network of a MIMD parallel
machine \cite {wha}. Later computations have been performed on an IBM SP/2 at the NASA Ames
Research Center and Thinking Machines Corp. CM-5 computer, and have consumed thousands hours of CPU
time \cite{wat}.
One of the main achievements of this study is a possibility to simulate
different motions (in this case high jumping and gait) using the same model.
The authors argue that development of high performance computers will make too
expensive the development of simple models for a specific motion and universal
models will dominate in the future \cite{and}.

A detailed neuromusckular model has been used by Ogihara \& Yamazaki
\cite{ogih}
to simulate human gait.
It consisted of 7 rigid links in the saggital plane with 9 muscles and a
sensory-motor nervous system.
The complete model includes 76 differential equations. Genetics algorithms
have been used to tune 93 parameters exploiting energy expenditure per unit
distance travelled as a goal function. The generated walking patterns were in
good agreement with
the observed ones.

\chapter {Conclusions}
 \begin{flushright}
{\small
This is not the end, it is not even the beginning of the end. But it is, perhaps, the end of beginning.\par
{\em Winston Churchill}
 \vspace{.3cm}
}
\end{flushright}

   The present brief review of the physically-based animations of human
motion shows that methods are mature enough to produce high quality
results for a large class of motions. However, there are a few problems that require additional study.

\section{Known Problems}

 \begin{flushright}
{\small
Complex problems have simple, easy-to-understand wrong answers. \par
{\em Murphy's Law Book Two}
}
\end{flushright}
 \vspace{.3cm}
The following list is probably incomplete and the order of issues is rather arbitrary.

\begin{itemize}
\item{{\em Soft tissue segmentation.} The development of an articulated model encounters a problem when one has to define
different soft tissue segments and  to describe sharp intersegmental boundaries.
This task is relatively easy for the head and the extremities. However, it is difficult
to cut for segments the neck, abdomen or thorax: these body parts are hardly can be called
soft tissue segments since they consist of a complex bone structure connected to various types
of soft tissue\footnote{For example, a reasonable division of the thorax into segments is practically
impossible (one has a choice: to consider the abdomino-thoracic complex as a single rigid body
or try to create a realistic thorax model that would include over three hundreds hard- and soft-tissue
subsegments).}.

\item{{\em Non-rigidity of the body segments.} Part of the body segments (muscle, connective tissues etc.)
can execute movements relative to the skeleton. Moreover, the properties of the relaxed and contracted
muscles are quite different. In \cite{hatze} a thesis by Gruber(1987) is cited. The author studied reaction
forces and moments in the knee and the hip during vertical-jump landing using
two three-link models: a rigid one and a model where "wobbling" masses had been attached to the rigid links.
She found that difference between the models could be up to a few thousand (!)
percent. Hatze  \cite{hatze} noted that these two models are the extreme ones
and in reality the muscle properties vary continuously.}

 \item{{\em Muscle models.} Majority of the dynamics-based simulations use a simple muscle model,
 in best case a three-element Hill's model. As has been already noted, more
accurate models are needed  to provide adequate results.}

\item{{\em An issue of choice of a goal function for optimization problems.}
As it was shown in \cite{bert}, a minimum metabolic cost of locomotion,
i.e. oxygen consumption per unit distance is a reasonable choice that provides
good results for a number of motions.
However, the authors themselves admit that agreement with experimental data
can be accidental and in reality quite different cost functions are used
by humans (for example, energy per unit time).
They also state that another cost function could be required for different
kind of motion (such as carrying loads, walking bent over under overhead
obstacles or walking with the aim to make no noise while hunting, for example).
In \cite{chou} it has been shown that humans appear to minimize energy expenditure when
           walking on level ground but not when stepping over obstacles:
           they use the larger-than-necessary obstacle clearance.}
}

\item{{\em Obstacle avoidance.} To the author's knowledge, up to date there is no
sufficiently general and fast method that will behave robustly, for example, for two highly
articulated figures' encounter
in some kind of fight.}



\item{{\em Computational cost.} The detailed models of human motion require
huge computational resources. Use of neural nets for emulation of real
physical processes \cite{nn2,nn1} is quite promising. Still, systematic
research to asses the optimal network architecture and learning algorithms is
needed. A special attention should be paid to hierarchical networks as the most
perspective ones for control of hierarchical articulated figures.}

\item{{\em Transient motions.} That seems to be one of the most difficult issues where
the motion capture
could not be of great help. If you have the motion of a running human and that of a lying, hardly
even intelligent procedures such as the neural nets trained on these examples will provide realistic
generalization for falling: evidently, in this transition some body parameters (masses and moments
of inertia) will play a crucial role while they are relatively unimportant in both running and lying
and thus cannot be learned by the neural nets.}
  \end{itemize}

\section{Physically-Based Simulation Engine for Animation}

\begin{flushright}
{\small
Wisdom consists of knowing when to avoid perfection.\par
{\em More Murphy's Law}
}
\end{flushright}

The features an ideal animation system should provide  are evident:

\begin{itemize}
\item{character's adherence to anatomy}
\item{character's adherence to physiology}
\item{physically accurate motion (if an animator does not specify a controllable exaggeration)}
\item{personification of the character}
\item{character's ability to master motor skills}
\item{character's interaction with environment and other characters}
\item{efficiency}
\item{compatibility of the data structure to that of rendering software}
\end{itemize}

This is certainly a toll order. Three less ambigous problems for which
physically-based simulation seems to be important are the following ones:
\begin{enumerate}
\item{Blending of captured motions}
\item{Retargetting of captured motion}
\item{Continuation of captured motion}
\end{enumerate}

It is worth to note that while the use of the underlying biology and physics is {\em desirable} for
the first two problems\footnote{it has been already stresses, however, that blending
of cardinally different motions by pure geometrical or signal
processing methods can give very poor results}, it is {\em mandatory}
for the third one if the character of motion is to be changed
as a result, for example, collision ("unintended" motion).
Retargetting can be considered in very broad sence: it can be an adjustment to a new character,
to a new emotional state of the old one or to environment.

\subsubsection{Implementation Tools}

  \begin{flushright}
{\small
The absence of alternatives clears the mind marvelously.\par
{\em Henry Kissenger}
}
 \vspace{.3cm}
\end{flushright}

To develop an efficient physically-based simulation core for animation system one should
use (wherever possible) and modify (wherever necessary) existing techniques.
There are a few guidance principles that should be respected:
\begin{itemize}
\item{hierarchical approach for}
{\begin{itemize}
\item{skeleton \& muscle representation}
\item{constraints}
\item{kinematics \& dynamics}
\item{control}
\end{itemize}}
\item{adaptivity (variable resolution) via use of}
{\begin{itemize}
\item{"passive" DOFs}
\item{reconstruction of missing DOFs}
\item{natural dynamics}
\end{itemize}}
\item{knowledge-based high-level motion control (domain-specific rules \& control parameters)}
\end{itemize}

Probably the most important tools for achieving the aims formulated are
multiresolution analysis (either B-spline or wavelet) and artificial neural nets\footnote{Neural nets used
for data analysis (with a few exceptions such as counterpropagation and self-orginizing
maps) are equivalent to  the well known statistical methods being usually much more slow. For example,
the most popular neural networks called multilayer perceptrons are just nonlinear regression and discriminant
models. Standard learning agorithms are inefficient because they have been designed for massively parallel
computers. On a serial computer neural nets can be trained more efficiently by standard optimization algotithms.}.
The former provides the decomposition of the motion into the basic one and "style" (personality)
and thus creates the basis for the motion transformations.

Neural nets could be used either as universal approximators (for example,
as efficient muscle \& dynamics models) or in data mining mode to extract hidden relationships
among motion parameters.

The most important application is probably for motion controllers generation:
an animator has no choice when dealing with unintended motion and has to use forward dynamics
based methods.  One should distingwish different kinds of motion depending on the role of motor control system:
kinematic tasks (slow motions in which dynamics effects are not significant), reflex-like behavior
("pre-computed" motions that can be modified on fly while maintaining some invariant
characteristics) and dynamic servoing (the laws of mechanics are to be integrated in this case).
A notion of primitive movements ("eigen-movements") that can be expressed as
stereotypical trajectories should be examined in two aspects: extraction of these
entities from captured or generated motions and their superposition/sequencing as
means to generate new motions (and first of all to provide reflex-like behavior).

  \vspace{.5cm}

A less formal conclusions can be found in Appendix.


 {\small
\bibliographystyle{plain}
\bibliography{rhm}
}

\chapter*{Appendix}

 \begin{flushright}
{\small
 It is NO use saying "we are doing our best". You have got to succseed in doing what is necessary.\par
  {\em Winston Churchill}
}
\end{flushright}
\vspace{1.5cm}

 \center{{\em \huge TheAnimator's  Eleven Commandments}}
\vspace{1cm}
{\em \bf
\begin{enumerate}
\item{Thou shalt never synthesize a motion that thou can capture}
\item{Thou shalt never mistake a physically correct motion for a believable one}
\item{Thou shalt never produce a periodic or symmetric motion}
\item{Thou shalt never copy the nature, but mimic it}
\item{Thou shalt exploit  hierarchical approach at all stages of animation}
\item{Thou shalt do the multiresolution analysis of  motion of thine and
remember that low frequencies are basic motion while high ones are style and personality}
\item{Thou shalt tell the kinematic movement from the dynamic one}
\item{Thou shalt let thy character to use natural dynamics and reflex movement whenever possible}
\item{Thou shalt never do forward dynamics without a proper controller, for
chaos and madness await thee at this path}
\item{Thou shalt override the automatically computed motion by manual input whenever thou are in doubt}
\item{Thou shalt abandon the hope to develop a universal animation method and
remember that large software projects are never finished, only released}
\end{enumerate}
}

\end{document}